\documentclass[mnsc,nonblindrev]{informs3_no_remark} 
\RequirePackage{tgtermes}
\RequirePackage{newtxtext}
\RequirePackage{newtxmath}
\RequirePackage{bm}
\RequirePackage{endnotes}

\OneAndAHalfSpacedXI 

\usepackage{natbib}
 \bibpunct[, ]{(}{)}{,}{a}{}{,}%
 \def\bibfont{\small}%

\usepackage{subfigure}
\usepackage{lscape,booktabs,longtable}
\usepackage{amsmath}
\usepackage[pdftex,colorlinks=true,urlcolor=blue,citecolor=blue,pdfstartview=FitH]{hyperref}
\usepackage{comment}

\usepackage{caption}
\usepackage[ruled,vlined]{algorithm2e}

\newenvironment{lyxlist}[1]
 {\begin{list}{}
  {\settowidth{\labelwidth}{#1}
   \setlength{\leftmargin}{\labelwidth}
   \addtolength{\leftmargin}{\labelsep}
   }}
 {\end{list}}

\TheoremsNumberedThrough     
\ECRepeatTheorems

\EquationsNumberedThrough    

\begin{document}


\RUNAUTHOR{Castellanos et al.}

\RUNTITLE{Silent Abandonment in Text-Based Contact Centers}

\TITLE{
Silent Abandonment in Text-Based Contact Centers: Identifying, Quantifying, and Mitigating its Operational Impacts}

\ARTICLEAUTHORS{%
\AUTHOR{Antonio Castellanos}
\AFF{Hebrew University of Jerusalem,
\EMAIL{antonio.catellanos@mail.huji.ac.il}}
\AUTHOR{Galit B. Yom-Tov, Yair Goldberg}
\AFF{Technion---Israel Institute of Technology,
\EMAIL{gality@technion.ac.il},
\EMAIL{yairgo@technion.ac.il}}
\AUTHOR{Jaeyoung Park}
\AFF{University of Central Florida,
\EMAIL{jaeyoung.park@ucf.edu}}
}


\ABSTRACT{
\textbf{Problem definition:} 
In the quest to improve services, companies offer customers the opportunity to interact with their agents using texting. Such contact centers face operational challenges distinct from traditional call centers because the measurement of common proxies for customer experience, such as customers' abandonment and willingness to wait (patience), are subject to information uncertainty.
A main source of such uncertainty is \emph{silent abandonment}. 
Silent-abandonment customers leave the system while waiting without notifying the system. 
As a result, the system is unaware that the customers left, and the agent's time is wasted. Furthermore, the system may not be aware (even retrospectively) whether a customer silently abandoned or was served.
Our goals are to estimate the magnitude of the silent-abandonment phenomenon and develop tools to mitigate its impact on the system.\\
\textbf{Methodology/results:} 
We develop classification models to investigate the magnitude of silent abandonment. 
A sample of 17 companies shows that 3\%--70\% of the customers entering these contact centers abandon silently. A detailed text-based analysis of one company identifies that 71.3\% of abandoning customers do so silently and that such behavior reduces agent efficiency by 3.2\% and system capacity by 15.3\%, incurring an estimated annual cost of \$5,457 per agent.  
Separate from this classification, we develop an expectation-maximization (EM) algorithm to accurately estimate customers' patience under information uncertainty and examine covariates influencing its value. \\
\textbf{Managerial implications:} 
We find that companies should use classification models to estimate abandonment scope and our EM algorithm to estimate customer patience. We suggest strategies to operationally mitigate the impact of silent abandonment by predicting suspected silent-abandonment behavior or changing service design. Specifically, we show that while allowing customers to write while waiting in the queue creates a missing data challenge, it also significantly increases patience and reduces service time, leading to reduced abandonment and lower staffing requirements.

}

\maketitle


%


\section{Introduction}
\label{sec:introduction}
The field of service engineering relies on measuring proxies for customer experience in a service system. Two of the most common operational measures used as such proxies are customer waiting and abandonment from the queue. Both are crucial performance measures for understanding customers' willingness to wait for service, which in turn is crucial for making operational decisions \citep{Mandelbaum2013Data,Garnett2002}. 
Waiting happens when a customer enters the service system, but there is not an available service agent. Abandonment naturally occurs when such waiting exceeds the customer's willingness to wait (henceforth, patience). Different streams of literature study various aspects of customer patience, such as its distribution (e.g., \citealt{Gans2003TelephoneProspects}), its connection to service utility (e.g., \citealt{Aksin2013}), and its manipulation (e.g., \citealt{Armony2008,Aksin2017}), among others.

The literature on estimating customer patience and its implications for optimizing operational decisions (e.g., staffing and routing) assumes accurate and complete knowledge of customer abandonment.
However, in some service environments, such as text-based contact centers (henceforth ``contact centers'' for short), we face the problem of not always being able to know whether a customer abandoned or received service, as we will explain shortly. This uncertainty creates a situation where the company is unsure of the service quality they provide to customers and how efficiently they use their resources, which in turn may lead to problematic operational decisions. The main goal of this paper is to overcome such uncertainties and allow companies to accurately estimate customer patience and abandonment by leveraging AI methods.

The abovementioned uncertainty is related to a customer behavior in contact centers queues which we term \emph{silent abandonment} (Sab).
A Sab customer is one that leaves the system while waiting in the queue but gives no explicit indication of doing so (i.e., they do not close the communication application when abandoning). Therefore, when a human agent becomes available, the (abandoning) customer is assigned to that agent. Only after the agent's inquiries go unanswered for some time does the agent (or system) realize that the customer has abandoned the queue without notifying the system, and the agent (or system) closes the conversation. We find that this situation creates two problems of {\bf information uncertainty}. The first is \textit{missing data}: the system may not be aware (even in retrospect) whether a customer silently abandoned the queue or was served (for a detailed definition of the concept of missing data, see \citealt{little2002statistical}). To the best of our knowledge, all companies assume the latter (that the customer was served), thereby inflating quality measurements upward.  The second problem is \textit{censored data}: the system may be aware that the customer silently abandoned the queue but it does not know exactly when, thereby censoring the data on customer patience  (for a discussion of censored data, see \citealt{smith_2002}). In addition, Sab customers create two operational problems of {\bf agent efficiency}:  \textit{partial idleness}, where the agent waits for inquiries from a customer that is no longer there and therefore some of their concurrency slots are idle; and \textit{wasted work}, where the agent tries to solve problems that have already been solved by the customer themself or by another agent (e.g., when the customer writes an inquiry while waiting in the queue, then abandons the queue and switches to a different communication channel, such as a phone call), thereby creating confusion, frustration, and wasted effort. 
We note that silent abandonment is more likely to occur when the system is overloaded with long queues and extended wait times. During such periods, a significant number of agents are likely to be ``busy" with abandoning customers, thereby wasting critically needed capacity.  Moreover, the Sab customers are taking the places of customers that want service and are actually waiting in the queue. 
Finally, we note that silent abandonment results in inaccurate measurements of queue length. Consequently, any algorithm that relies on this information (e.g., for  delay announcements; see \citealt{Ibrahim2018DelayAnnReview}) must be adjusted to account for silent abandonment.

The abovementioned missing data problem is connected to the company's inability to  distinguish between Sab and served customers even retrospectively without reading each conversation's text. 
Specifically, text-based contact centers usually allow customers to communicate their problems while waiting in the queue, 
creating uncertainty regarding customers who wrote while waiting but did not write anything after an agent was assigned to serve them. Without reading the conversation text, it is impossible to distinguish whether such customers have abandoned the queue silently or were served but did not acknowledge being served, even by a thank-you message.
We refer to a customer from this class as an \textit{uncertain silent abandonment} (uSab) customer and build classification models to separate them to their respective customer groups (see Section \ref{sec:dataAndReseatch}). 

One might think that preventing customers from writing messages can eliminate the system problem. We note that while such policy will certainly alleviate the problem of missing data, it will not solve the censoring or the agent efficiency problems mentioned above. Furthermore, we show in Section \ref{sec:EMextension-Model} that customer ability to write while waiting plays an important factor in shaping their experience in the system and triples their willingness to wait. 
It is also important to note that the industry trend is to allow such freedom to customers. In fact, our industry partner reports that by 2020, 66\% of their clients---including 35\% of \textit{Fortune}'s most valuable companies---had adopted systems that allow such operation, a significant increase from just 5\% in 2016 \citep{LivePerson2020}. This rapid adoption underscores the growing prevalence of this challenge and highlights the importance of addressing it.

The phenomenon of Sab is very common in contact centers, as we demonstrate in Section \ref{sec:dataAndReseatch}.  
Using a sample of companies from different domains and sizes,
we show that the proportion of uSab customers ranges between 3\% and 70\% of the contact centers' conversations (22\% on average across companies). But these percentages serve only as an \emph{upper bound} to the Sab proportion since some uSab customers were served; hence, inferring how many of them indeed abandoned is important. In Section \ref{sec:prob_abnd_WQ}, we analyze in detail one contact-center database 
where the uSab percentage is 24.4\%---close to the industry average. Using our classification model, we conclude that 51.8\% of these uSab conversations are indeed silent abandonment (12.6\% of the contact-center conversations), while the rest are classified as served. Furthermore, we show that Sab represents 71.3\% of the total abandonment in the system (see Table \ref{tbl:general_stat}). With all the above, we see that silent abandonment is a widespread phenomenon requiring careful consideration to understand and manage.
sub

\subsection{Paper Structure and Summary of Contributions} \label{sec:goals}
\begin{enumerate}

\item \textit{Estimate the scope of the silent-abandonment phenomenon}. In Section \ref{sec:dataAndReseatch}, we identify and estimate the proportion of customers who silently abandon the queue in contact centers.  
Initially, we classify uSab customers based on the number and timing of their messages, providing an upper bound for the Sab proportion. In Section \ref{sec:prob_abnd_WQ}, we refine this classification by constructing a model to subclassify uSab customers to their respective classes, enabling a detailed estimation of silent abandonment from the individual customer level to the system level.
The classification model leverages multiple features, including the text messages of the customer and agent. Applying the model to the XYZ dataset reveals that approximately two-thirds of abandoning customers silently abandon. 

\item  \textit{Create an algorithm to estimate customer patience in the presence of silent abandonment}. 
As we mentioned, customer behavior in contact centers differs from that in call centers, creating two types of uncertainty in the data (i.e., censored data and missing data). To our knowledge, no paper has attempted to estimate customer patience in contact centers, although finite patience has been considered in optimization models of contact centers  \citep{Tezcan2014RoutingCustomers}.  
In Section \ref{sec:patience_estimate}, we develop a new method for estimating customer patience addressing both uncertainty types. This method uses an expectation-maximization (EM) algorithm. We show that the EM model provides a more accurate estimation than does classifying conversations and using them with classical estimation methods. We observe that customer patience in the XYZ contact center is very long---81 minutes---and discuss possible explanations for it.

\item  \textit{Understand factors associated with longer patience.} 
In Section \ref{sec:patience_coef}, we argue that not allowing the customer to write in the queue may reduce ambiguity, offering the advantage of more accurate patience estimation, but it does not eliminate the Sab phenomenon. 
Moreover, we claim that such a policy unnecessarily limits the customers' ability to express themselves, reduces customer patience, and wastes agent capacity. It also limits the company’s ability to make informed decisions about customer prioritization. 
To support these claims, in Section \ref{sec:EMextension-Model} we extend our EM patience estimation algorithm to account for various factors that may influence customer patience. This algorithm allows us to estimate patience according to customer activity in the queue. In Section \ref{sec:EMextention-Patience Estimation}, we show that customers who write while waiting exhibit 2.83-times longer patience. Furthermore, in Section \ref{sec:patience_coef}, we show that customer writing while waiting reduces service time by 13.2\%. Therefore, preventing customers from writing will come at the cost of reduced efficiency, and companies should use other policies to reduce the burden of Sab. We suggest such policies in Sections \ref{sec:avoid_Sab} and \ref{sec:BoostingExpiriance}.

\end{enumerate}

\section{Literature Review}
\subsection{Context}
The context of this research is contact centers, which are an important part of the service industry's digital revolution. Service companies branch into more accessible and easy-to-use service channels such as mobile applications. Technology allows modern-day companies to replace traditional service encounters (face-to-face, telephone) with technology-mediated ones \citep{Massad2006CustomerEncounters,vanDolen2002ModeratedEncounter}, which allow customers and service employees  to connect through digital interfaces \citep{Schumann2012TechnologyAcademics,Froehle2004NewExperience}. 
Today, employees and customers can interact through social media (e.g., $\mathbb{X}$ or Facebook), chats on corporate websites, or messaging applications (e.g., WhatsApp or WeChat). This enables customers to interact with agents through platforms similar to those used to contact their family and friends. 
Unsurprisingly, text-based contact centers are gradually replacing call centers as the preferred way for customers to communicate with companies. Indeed, as early as 2012, a survey conducted by a contact-center solution provider found that 78\% of customers preferred texting with a company to calling its call center \citep{ringCentral_2012}.

This digital revolution provides the service industry not only new opportunities to improve services \citep{Rafaeli2017TheFuture,Altman2021} but also new operational challenges. 
Operating text-based contact centers is substantially different from operating call centers. For example, in textual service systems, unlike in call centers or face-to-face services, agents can provide service to multiple customers concurrently \citep{Tezcan2014RoutingCustomers,Goes2017WhenMultitasking,daw2021coproduction}. Furthermore, the phenomenon of silent abandonment, researched here for the first time, is the result of such text-based services. 


\subsection{Scope of Silent Abandonment in Different Contexts}
Sab of customers waiting in a queue is not exclusive to contact centers and appears in other environments such as ticket queues \citep{Kerner2017,Kuzu2019ToWait}, scheduling systems \citep{Liu2016},  emergency departments (EDs) \citep{Yefenof2018}, and interactive voice response (IVR) systems \citep{Carmeli2019IVR}. 

In ticket queues, an arriving customer receives a ticket with their queue number (and possibly estimated wait time) \citep{Kerner2017}. The customer may silently abandon either before joining the queue or at some point during the wait. Analogous to contact centers, the Sab is not realized until the customer is called for service and does not show up, and the exact abandonment time is unknown. \cite{Kuzu2019ToWait} estimate the scope of abandonment in ticket queues of a bank to be 13.2\%--26.1\%, depending on the customer class. High abandonment rates in ticket queues indicate inefficiencies in the service delivery of the company, potentially damaging revenue, customer satisfaction, and loyalty \citep{Kuzu2019ToWait}. 



Silent abandonment is also conceptually similar to no-show incidents, 
where a customer fails to arrive for a scheduled appointment without prior notification. 
No-show incidents are particularly common in healthcare, with rates ranging from 23\%--34\% \citep{Liu2016}. Healthcare providers can reduce this phenomenon by reminding the customer of their future appointment \citep{Geraghty2008}, but cannot eliminate it. \cite{HoLauNoshows1992} showed that no-shows strongly affect system performance because of capacity loss and forced physician idleness. We claim these efficiency problems also happen in contact centers due to Sab. 
Overbooking and appointment-book design policies are used to reduce the idleness resulting from no-shows by controlling the number of customers arriving at each time slot \citep{Vissers1979}. Then, when a no-show is detected, the agent is able to serve another customer without delay. We cannot apply such methodologies in contact centers, where arrivals are unscheduled, but an analogy can be made to concurrency policies, where an agent may serve other customers if one does not react. We suggest in Sections \ref{sec:avoid_Sab} and \ref{sec:BoostingExpiriance} methods to reduce the adverse impact of Sab on efficiency by adjusting concurrency polices and more.  

Silent abandonments also occur in emergency departments, where patients may abandon the queue before seeing a medical practitioner without telling anyone---a phenomenon called left without being seen (LWBS). According to \cite{Medicare}, the national average of LWBS patients in US EDs was 2\% during 2021 (much lower than Sab in contact centers). ED abandonment increases the risk of a patient suffering an adverse outcome, increases the probability of the patient returning to the hospital \citep{Baker1991}, and impacts hospital revenue \citep{Batt2015}. 
\cite{Improta2022Abandonment} use machine learning approaches to predict LWBS in real time. Here too, we use AI to classify Sab, and suggest that such models could be used to identify suspected Sab customers in real time (see Appendix \ref{app:ClassModQueue}).

In all the above contexts (ticket queues, scheduling systems, and EDs), the system knows retrospectively whether the customer abandoned or showed up. In contact-center systems, this is not so; the conversation metadata (i.e., operational data and timestamps) does not distinguish between served and abandoned customers. We therefore use text-analysis classification models to classify customers and estimate the Sab scope, based on the conversation text (see Section \ref{sec:prob_abnd_WQ}). 

IVR systems also have a similar classification problem. There customers terminate the navigation process and leave the system, with no clear indication whether they abandoned or were served by the IVR. 
\citep{Carmeli2019IVR} estimated the proportion of customers abandoning the IVR as 17\% by fitting a mixture distribution to IVR service durations, relying on prior information of the IVR process structure. They discuss how abandonment measurement can help companies to better design their IVR systems. 


\subsection{Estimating Customer Patience}
While silent abandonment manifests across various environments, a deeper understanding of its operational impact requires robust methods for customer-patience estimation. Estimating patience is critical to predicting abandonment behaviors and designing effective service strategies to mitigate Sab. The literature on estimating patience has primarily focused on two environments: call centers and emergency departments. 
Many studies 
have developed methods to estimate customer patience in call centers, where service dynamics create right-censored data (see review \citeauthor{dai2011queues} \citeyear{dai2011queues} and references therein as well as  \citeauthor{Aksin2017} \citeyear{Aksin2017}, \citeauthor{emadi2018customer} \citeyear{emadi2018customer}, \citeauthor{Ye2020HazardRate} \citeyear{Ye2020HazardRate}). 
For example, \cite{Mandelbaum2013Data} assumed that customer-patience time, $\tau$, and virtual wait time, $W$, are exponentially distributed with rates $\theta$ and $\gamma$, respectively, and developed a maximum likelihood estimator for customer patience from right-censored data. 
In contrast, EDs face the additional challenge of left-censored data due to the phenomenon of LWBS. 
Specifically, \cite{Yefenof2018}---an important predecessor to this paper---developed maximum likelihood methods to address data censoring when estimating patient patience in EDs.
Yet, all the above assumes complete data, whereas our EM model takes into account the class uncertainty between Sab and being served.

The literature on patience also investigates what affects customer patience, showing, for example, that delay announcements \citep{Kerner2017}, system load \citep{Ye2020HazardRate,Bolandifar2023}, patient acuity \citep{Bolandifar2023}, and past waiting experience \citep{emadi2018customer} shape customer patience. We add to that literature the impact of service initiation, showing that customers who write their inquiries while waiting exhibit longer patience. The abovementioned papers \citep{emadi2018customer,Ye2020HazardRate,Bolandifar2023} had the advantage of complete information that enabled the use of hazard models in such investigation.  Here, we have missing information where conversation classification as abandonment is uncertain; therefore, we developed an EM model with covariates (\S\ref{sec:EMextension-Model}) to overcome this challenge. 


\section{The Scope of Silent Abandonment in Contact Centers and Its Impact on Efficiency} \label{sec:dataAndReseatch}

We posit that the phenomenon of silent abandonment (Sab) is pervasive in contact centers. To substantiate this claim, we collected general metadata from several Western companies and obtained detailed data from a telecommunications company in the United States, which we refer to as ``XYZ Company." These datasets were provided by LivePerson Inc., a leading provider of computational infrastructure for the contact-center industry.

In this section, we first describe the dynamics of conversations in contact centers and highlight the challenges associated with estimating abandonment probabilities. We demonstrate how customer behavior significantly influences informational uncertainty. Subsequently, we estimate the true abandonment probability by uncovering the prevalence of silent abandonment in our data using an AI-based classification model applied to the XYZ dataset.

\subsection{The Service Process and the Data Describing It}
\label{subsec:MessagingData}

\begin{figure}[tbh]
\centering
\subfigure[Sr Customer]{
\includegraphics[width=0.22\textwidth]{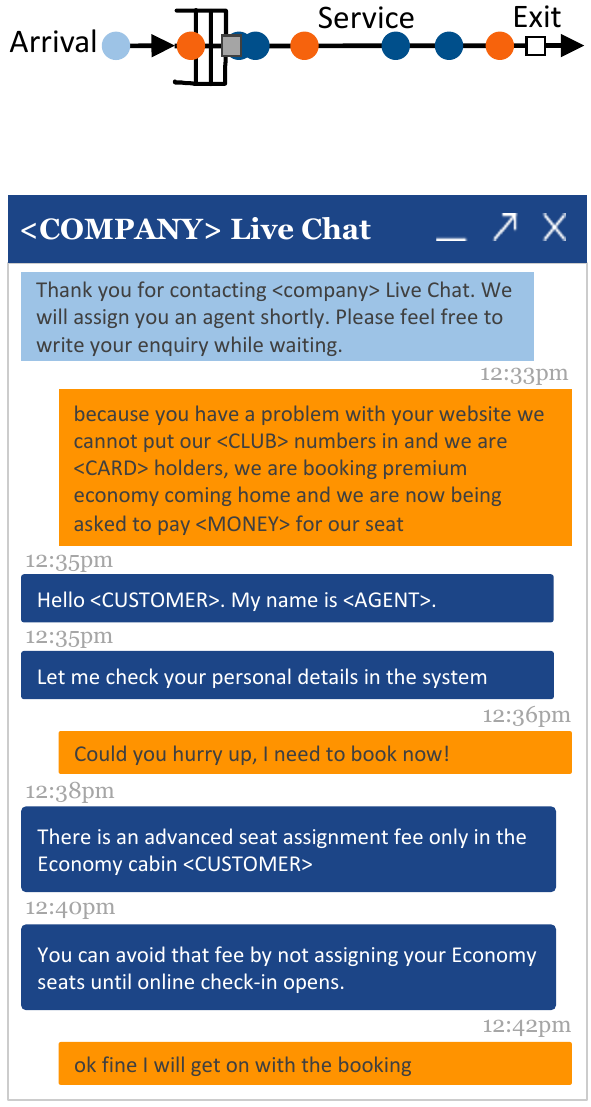} \label{fig:SRW}
} 
\subfigure[Kab Customer]{
\includegraphics[width=0.22\textwidth]{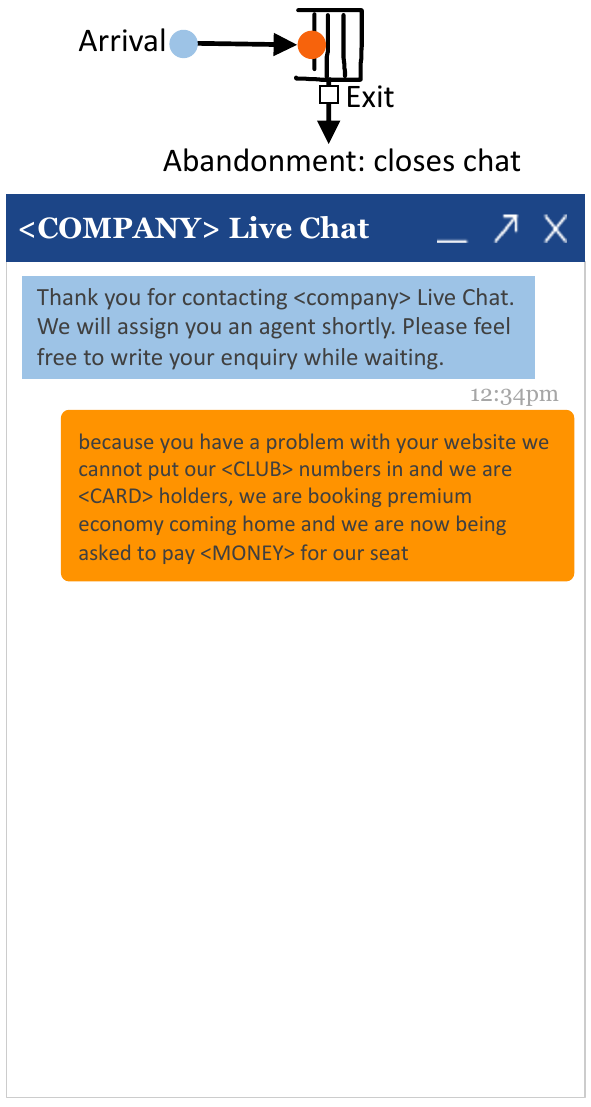} \label{fig:KABW}
} 
\subfigure[uSab: Sr1 Customer]{
\includegraphics[width=0.22\textwidth]{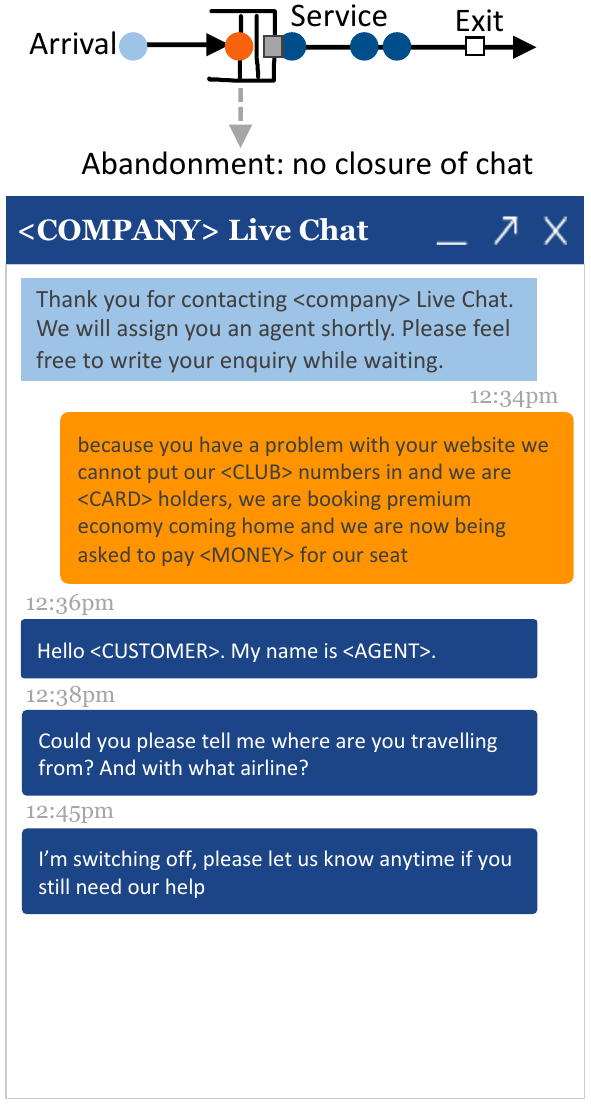} \label{fig:ShortServConvo}
} 
\subfigure[uSab: Sab Customer]{
\includegraphics[width=0.22\textwidth]{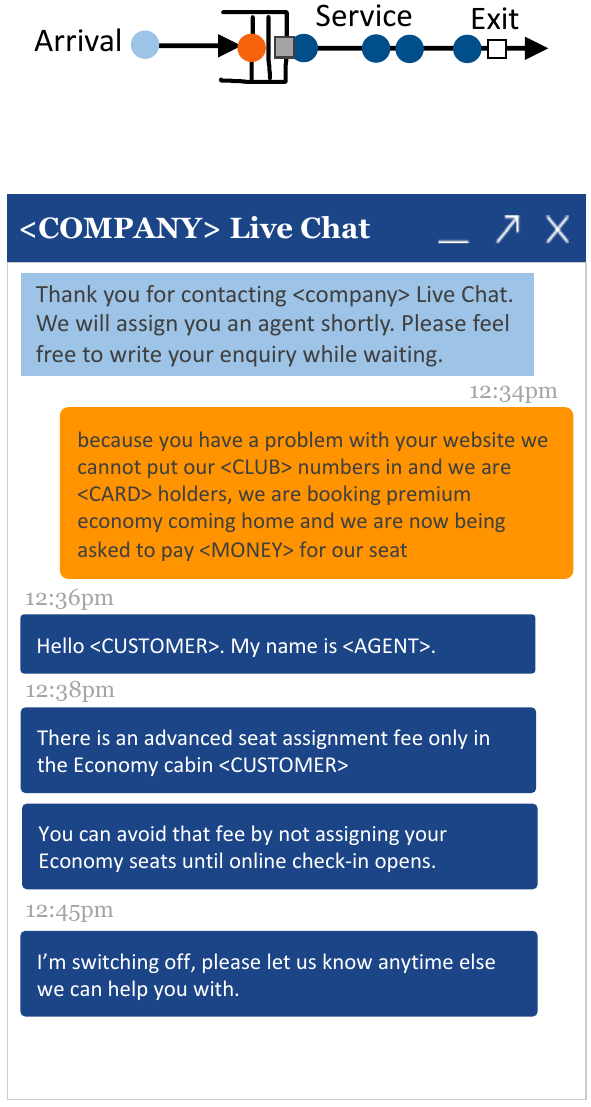}
\label{fig:SabConvo}
}
\caption{Example Conversation and the Metadata Process Flow of the Four Types of Customers in Contact Centers}
\label{fig:cust_Process_flow_WQ}
\end{figure}

To explain how contact centers operate, we stylize several examples of conversations, each representing a distinct service outcome. These conversations are presented in Figure \ref{fig:cust_Process_flow_WQ}, where agent messages are depicted in blue and customer messages in orange. The figure outlines four scenarios of conversation dynamics:
\begin{enumerate}
    \item \textbf{Served Customer (Sr)}:
    These customers enter the queue, wait, are assigned to an agent, and while served engage in a back-and-forth exchange with the agent. Customers may write their inquiry either while waiting or after being assigned to an agent.
    \item \textbf{Known Abandonment (Kab)}: These customers enter the system, wait, and at some point decide to close the communication app, signaling to the system that they have abandoned the service. The system records their abandonment and its timestamp, ensuring they are not assigned to an agent. Customers in this category may write an inquiry while waiting.
    \item \textbf{Served with One Exchange (Sr1)}: These customers enter the queue, wait, and write an inquiry while waiting. They are then assigned to an agent who responds. However, the service interaction consists solely of agent messages, with no further communication from the customer.
    \item \textbf{Silent Abandonment (Sab)}: These customers enter the queue, wait, and write an inquiry while waiting. They abandon the service app without explicitly closing it. Consequently, they are later assigned to an agent, but the agent is unable to serve them as they are no longer active in the system.  
\end{enumerate}

The metadata of a conversation includes timestamps for customer arrivals, timestamps of messages written by either the customer or the agent, and the number of customers each agent served concurrently at any given time. (It is important to note that in the contact center we studied, interactions involve only human agents, not bots.) Additionally, the metadata records information about who closed the conversation---whether it was the customer, the agent, or the system. In cases where the system automatically closes the conversation, it does so after a predetermined period of customer inactivity (e.g., two hours in the XYZ contact center, referred to as the ``automatic closure time").

Note that the last two customer categories described above (Sr1 and Sab) appear identical from a metadata perspective. Both involve customers who wrote an initial query while waiting, were assigned to an agent, but did not send any subsequent messages. Distinguishing between these two groups requires analyzing the text of the conversation. For this reason, we refer to the unified group of customers who exhibit this behavior as \textbf{uncertain silent abandonment (uSab)} customers.
The challenges of classifying these uSab customers and overcoming the missing information regarding the exact timing of abandonment within the Sab subgroup lie at the core of the issues addressed in this study.

\begin{figure}[htb]
\centering
\includegraphics[width=0.40\textwidth]{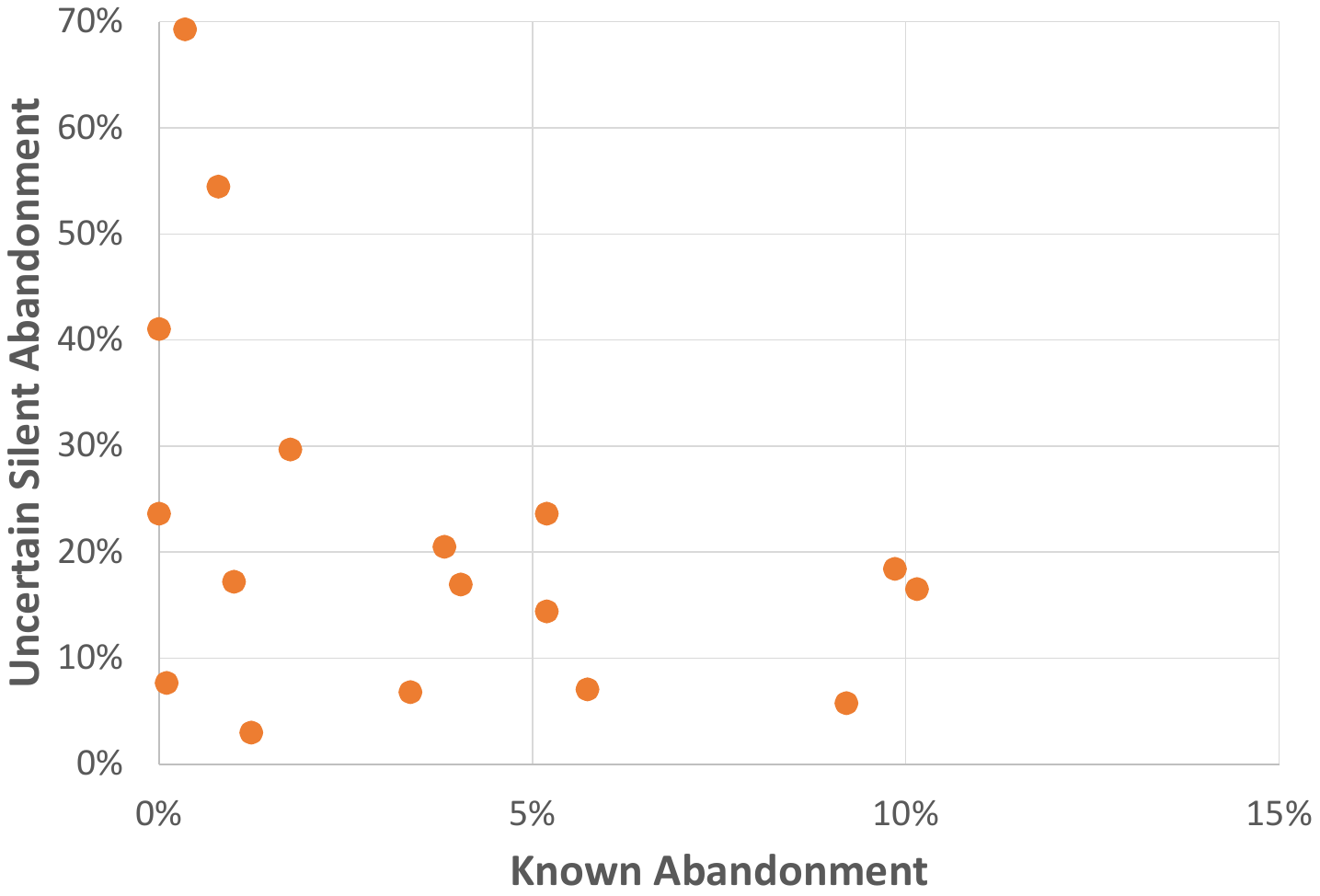}%
\caption{Proportion of Known  and Uncertain Silent Abandonment in Different Contact Centers}
\label{fig:Sabfig}
\end{figure}

Next, we analyzed metadata from 17 Western companies spanning various domains and sizes. These companies range from small contact centers handling approximately 23,000 conversations per month to Fortune 500 firms managing over 740,000 conversations monthly on average. For each company, we calculated the proportions of Kab customers and uSab customers entering the contact center. These proportions are presented in Figure \ref{fig:Sabfig}.
The proportion of uSab customers across these 
companies ranges from 3\% to 70\%, significantly higher than the Kab proportion, which falls between 0.1\% and 10.1\%. However, the uSab proportion serves only as an \emph{upper bound} for the Sab proportion, as some uSab customers ultimately receive service. On average, uSab customers account for 22\% of all conversations. Therefore, accurately determining how many of these customers truly abandoned is crucial for understanding the true service level the companies provide.


To further investigate the Sab phenomenon, we obtained detailed data from XYZ Company, encompassing 332,978 service interactions conducted in May 2017. During data cleaning, 1,391 conversations (0.4\%) were excluded due to unrealistic or incomplete information. The dataset includes comprehensive metadata (as described above) for all conversations as well as the text written for uSab conversations.

The XYZ contact center operates 24/7, with arrivals varying throughout the day. On average, there are 594.79 customer arrivals per hour. The mean number of online agents per hour is 134.69, with agents handling an average of 3.98 customers concurrently ($SD=4$).

The average total time a customer spends in the system (from entry to conversation closure) is 120.40 minutes ($SD=95.80$) (an average that includes Sab customers).  
The average wait time in the queue is 8.30 minutes ($SD=18.45$), while the average service duration from agent assignment to the last message is 46.34 minutes ($SD=63.56$). 
On average, conversations remain open for 65.93 minutes ($SD=75.08$) after service concludes---this time is influenced by the automatic closure time of two hours that XYZ Company applies.

We analyzed customer behavior in the queue prior to assignment to a service agent to investigate the relationship between in-queue behavior and abandonment. On average, customers sent 1.20 messages while waiting ($SD=0.51$). The initial messages contained an average of 105.60 characters ($SD=125.86$) and 18.17 words ($SD=24.64$).

Further details about the XYZ dataset are provided in Appendix \ref{app:summary_stats}. General statistics for the conversations, such as the number of messages, are summarized in Table \ref{tbl:general_stat}, which includes data for all conversations and a breakdown by customer type. Additionally, the distribution of conversation start days and hours is illustrated in Figure \ref{fig:ChatStart}.

\subsection{Estimating the Scope of Silent Abandonment}
\label{sec:prob_abnd_WQ} 
At the XYZ contact center, 5.1\% of customers abandoned the queue by explicitly closing the communication app, classifying them as Kab. uSab customers represent 24.4\% of all conversations, while the remaining 70.5\% of customers clearly received service and are categorized as Sr.

In this section, we develop a classification model to distinguish uSab conversations into the Sr1 and Sab categories. This classification enables us to accurately estimate the true proportion of abandoned customers and mitigate bias introduced by the exclusion of Sab customers. Furthermore, the model helps estimate the time required for a service agent to recognize that a Sab customer has abandoned the queue---time that constitutes wasted effort. We further quantify the extent of the agent's wasted effort on such interactions.

We propose a detailed examination of conversation transcripts to classify whether uSab customers silently abandoned the queue or were served. 
To this end, we manually labeled a random sample of 650 uSab conversations into two groups---Sr1 and Sab---by reviewing the full conversation transcripts. The sample comprised 342 Sr1 conversations and 308 Sab conversations.

Next, we extracted textual features from the conversation transcripts and metadata features described in Section \ref{subsec:MessagingData}. To derive textual features, we employed natural language processing (NLP) techniques. 
We built a sparse matrix by tokenizing the transcript of the conversations and filtering words that do not convey information (like ``a'' and ``the'') using the English stop words dictionary in scikit-learn \citep{scikit-learn}. 
For each word in the sparse matrix, we computed its mutual information \citep{kraskov2004estimating}, which measures dependencies between attributes, to assess how much information each word contributes to the silent-abandonment tag. (This results in a mutual information matrix.)
To reduce dimensionality, we selected the top 50 agent words and top 50 customer words with the highest dependency on the Sab tag according to the mutual information matrix. Each selected word was then represented by a variable indicating its frequency of occurrence in the conversation.

For model development, we used a random subset of 550 conversations, reserving the remaining 100 conversations for a final out-of-sample evaluation. The 550-conversation set was randomly divided into training and test sets, containing 75\% and 25\% of the conversations, respectively.

Let $\pi_{i}$ denote the probability that customer $i$ silently abandoned the queue, given that this conversation is part of the uSab group. Formally, $\pi_{i} \triangleq \textit{Pr}\left\{ Sab_{i}\mid uSab_i\right\}$.

\begin{table}[!htb]
   \begin{minipage}{.5\textwidth}
    \centering
    \includegraphics[width=0.75\textwidth]{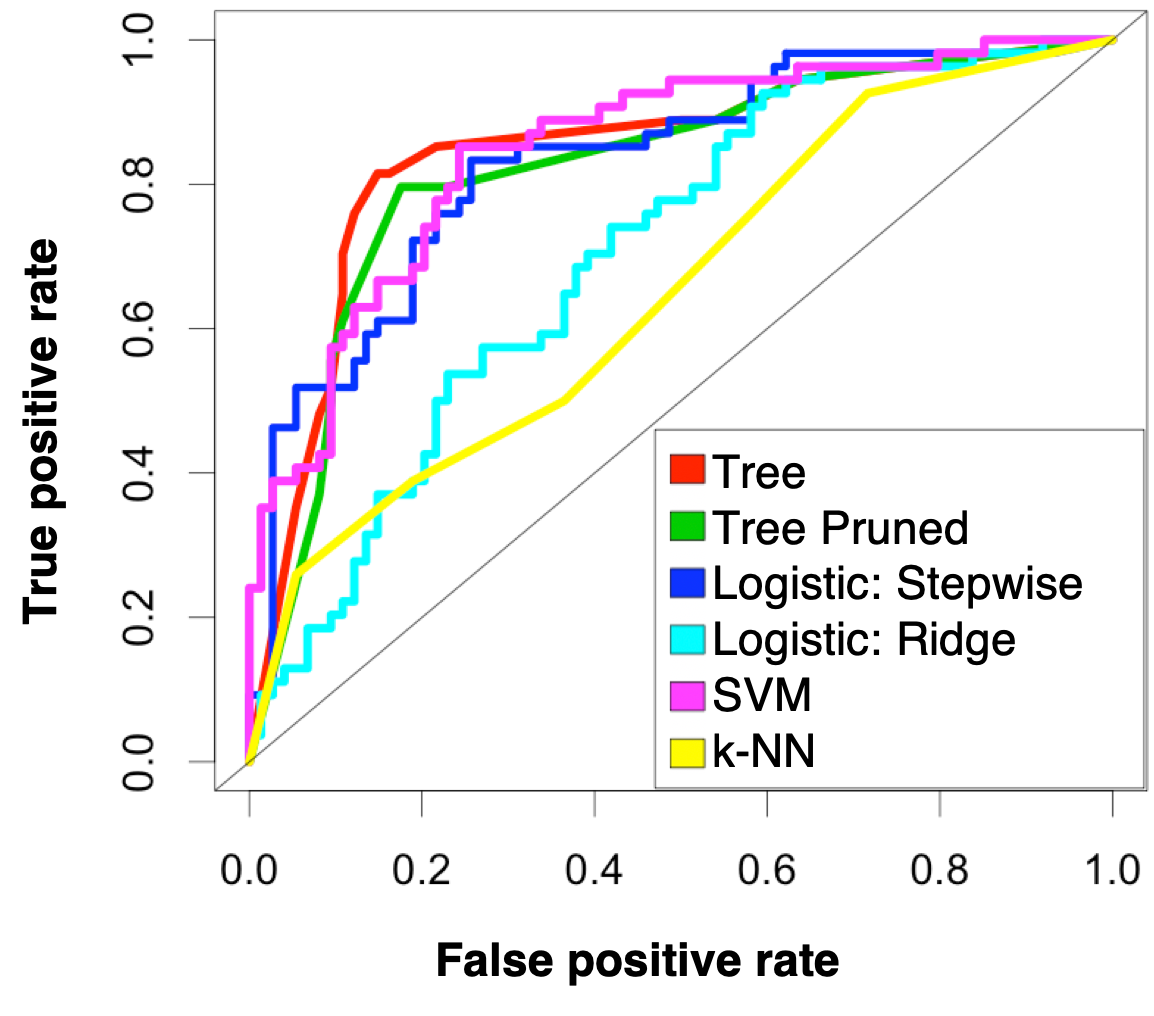}%
    \captionof{figure}{ROC Curve on Test Dataset}
    \label{fig:rocC}
  \end{minipage}
  \begin{minipage}{0.3\textwidth}
    \caption{Area under the ROC}  
    \label{tbl:logridge}
    \centering
    \begin{small}
    \centering
    \begin{tabular}{lc} 
    \toprule
    Model & AUC\\ \midrule 
    SVM & 0.85\\
    Tree & 0.85\\
    Logistic Regression: Stepwise & 0.83\\
    Tree Pruned & 0.82\\
    Logistic Regression: Ridge & 0.71\\
    k-NN & 0.65\\     
    \bottomrule 
    \end{tabular}
    \end{small}
  \end{minipage}\hfill
\end{table} 

We examined the performance of several AI classification methods: logistic regression (stepwise backward and with a ridge penalty), support vector machines (SVM), k-nearest neighbors (k-NN), and classification tree (additionally, we pruned the tree). We trained each model on the training subset of 412 conversations defined above, and then estimated $\hat{\pi}_i$ for each conversation $i$ in the test subset (138 conversations).
In Figure \ref{fig:rocC}, we compared the accuracy of these classification models using the receiver operating characteristic (ROC) curve. The ROC plots the true positive rate (TPR) against the false positive rate (FPR) for varying threshold levels. The ROC curve is a recognized visual method for comparing performance of different classification methods and for selecting the best threshold to work with \citep{Fawcett2006ROC}. A standard characteristic in that regard is the area under the ROC curve (AUC), presented in Table \ref{tbl:logridge}.   Using this criterion, we conclude that the best classification methods for our problem, with $\text{AUC} = 0.85$, are the SVM model and the classification tree. Models with an AUC above 0.80 are considered ``excellent" classification models \citep {Hosmer2000}. More details about feature selection procedures used for the SVM and the classification tree models can be found in Appendix \ref{app:ClassMod}. The SVM model includes the following features, among others: specific words written in the conversation, customer experience (e.g., amount of time the customer waited in the queue), and agent's work time (e.g., amount of time the agent engaged with the customer). The SVM model includes words showing whether the customer is seeking the agent's immediate attention along with specific words related to the service the company provides. Due to privacy and legal constraints, we cannot share the exact expressions. However, we can summarize that Sab customers typically expressed a sense of urgency in obtaining a resolution. 

Our chosen model is the SVM. To select a specific threshold level for the model, we searched for the threshold that maximizes the sensitivity (TPR) and specificity (1-FPR) proportions; that is, it maximizes the proportion of Sab and Sr1 conversations that are correctly identified. To that end, we use the Youden index, defined as $\max_{c}\{\mathrm{sensitivity\left(c\right)+specificity\left(c\right)-1}\}$, where $c$ is the threshold \citep{berrar_2019}. Note that in our setting, this optimization problem is not weighted, since we care as much about maximizing the proportion of Sab correctly identified as we do the proportion of Sr1 customers. 
We find that the optimal threshold ($c$) is 0.47 with a sensitivity proportion (TPR) of 85\%, a specificity proportion (1-FPR) of 76\%, and an error rate\footnote{Error rate is computed using the false positive rate and the false negative rate as follows: $\left(\frac{FPR+FNR}{N}\right)$, where $N$ is the sample size.} of 20\%. An additional way to choose the best threshold is by maximizing the F1-score (a harmonic mean of precision and TPR) \citep{Chinchor1992}, that is, by maximizing the sum of precision and recall together. The highest F1 score in our case is 0.78 obtained with the same threshold $c=0.47$.
We then performed a validation test on an out-of-sample set defined above (including 100 random uSab conversations separate from the 550 conversations used for the model and threshold selection). We find that 91\% of the out-of-sample conversations are correctly classified by the SVM with that specific threshold (0.47). 

To obtain results for the full write-in-queue dataset, we processed the transcript of every uSab conversation and indicated whether it contained the SVM's words. We classify each conversation as Sab if the $\hat{\pi_{i}}$ estimated by the SVM is larger than 0.47. 
We find that out of the group of uSab conversations, which constitute 24.4\% of all the conversations in the data, 51.8\% are classified as Sab conversations (have an SVM score larger than 0.47) and 48.2\% are Sr1 conversations. This means that the actual proportion of abandoning customers in this dataset is 17.7\%, far above the estimation of 5.1\% abandonment that the company currently has based on the Kab conversations. Moreover, out of all abandoning customers, we find that 71.3\% are Sab customers (12.6\% of all arriving customers). This again highlights the importance of taking Sab customers into account in order to correctly evaluate performance levels in contact centers. 

To estimate the ``service time" of customers that silently abandoned the queue, we calculate the average service duration of the Sab conversations classified above (using the $\hat{\pi_{i}}> 0.47$ threshold). We find that such conversations take, on average, 20.06 minutes ($SD=40.65$), from assignment until last message. One of the reasons for this long duration is company service policy. According to the company policy, an agent must reach out to an unresponsive customer at least twice (after the initial message) with at least 10 minutes between each reach (see Table \ref{tbl:general_stat} for statistics regarding within-conversation dynamics). While the agent might be serving other customers concurrently during that period, they need to be attentive to the passage of time to apply this policy and write the required messages. Note that during these 20.06 minutes, concurrency slot capacity is wasted while the agent tries to communicate with a departed customer. Given the uSab conversation classification, we estimate that the average service duration of Sr1 conversations is 53.67 minutes ($SD=71.36$). This finding reveals that on average Sr1 customers have a longer service time than Sab customers do and a similar service time to the Sr customers, which is not easily observed in the distribution of the length of stay (LOS) of uSab.\footnote{We note that the dynamics of Sr1 and Sr conversations  are different, which results from the abovementioned policy of how to handle unresponsive customers. Specifically, the data shows that the agent sends an Sr1 customer a mean of 4.31 $(SD=2.45)$ messages before ``giving-up" on them and either closes the conversation or lets the system close it automatically 2 hours later. On the other hand, during Sr conversations, when the customer is responsive, the agent sends  a mean of 11.8 $(SD=9.36)$ messages but does not need to wait the entire 10-minute interval, and therefore the time between messages is much shorter---2.99 minutes on average throughout the conversation (see Table \ref{tbl:general_stat}).} 
Moreover, we note that Sab customers have a longer average closure time, 113.64 minutes ($SD=65.88$), than Sr1 customers do, 94.98 minutes ($SD=74.99$). 

Using the SVM model and the threshold of $\hat{\pi}_i > 0.47$ mentioned earlier, we can also estimate the true wait times and quantify how these measures are biased by silent abandonment (Sab). For instance, XYZ Company estimates that their abandoning customers wait 7.3 minutes on average ($\mathbb{E}[W \mid \text{Kab}]$; see Table \ref{tbl:general_stat} for relevant statistics). However, our analysis reveals that these customers actually wait much longer---15.9 minutes on average ($\mathbb{E}[W \mid \text{Kab}\cup\text{Sab}]$). Conversely, the company believes that served customers wait on average 8.9 minutes ($\mathbb{E}[W \mid \text{Sr} \cup u\text{Sab}]$), whereas their actual wait time is shorter ($\mathbb{E}[W \mid \text{Sr} \cup \text{Sr1}]$). These discrepancies have implications for operational decision-making. Specifically, while served customers receive slightly better service than the company believes, abandoning customers experience far worse wait times. Correcting these misconceptions is crucial for ensuring that decisions are based on accurate insights into customer experiences.

We find that 13.3\% of the inquiries the agent answers are from Sab customers, 12.4\% from Sr1 customers, and 74.3\% from (regular) served customers. Measuring agent effort in treating Sab customers,
we find that agents waste on average 2.29 messages ($SD=1.29$) and 26.2 words per message ($SD=20.29$) per Sab customer. Taking an overall perspective, this comprises 3.2\% (3.6\%) of messages (words) written by the agents in that contact center.
%
To provide an additional estimation of wasted effort on Sab customers from the system perspective, we compute the amount of agents' concurrency time wasted on those customers.  Dividing the concurrency time spent on Sab conversations by the total concurrency time reveals that the system spends 15.3\% of its agents' concurrency capacity dealing with Sab conversations:
\begin{equation}
\label{eq:Effort_msg}
    \text{Time effort} = \frac{0.133\cdot(20.06+113.64)}{0.133\cdot(20.06+113.64)+0.124\cdot(53.67+94.98)+0.743\cdot(48.57+59.18)} = 0.153.
\end{equation}
We can go further by translating capacity loss into monetary terms. In the U.S., a chat-contact-center agent earns an average of \$35,664 annually \citep{ZipRecruiter2024}. This means that companies lose approximately \$5,457 per agent each year due to silent abandonment.  These findings highlight the importance of promptly identifying silent abandonment to enhance system efficiency and minimize financial losses. 

\subsection{Discussion: Reducing the Operational Impact of Silent-Abandonment Customers}
\label{sec:avoid_Sab}
We propose three strategies for companies to reduce the Sab-related capacity loss: 
\begin{enumerate}
    \item \textbf{Deploying bots} to handle \emph{suspected} Sab customers (e.g., over the 0.47 threshold in our SVM model). 
    \item \textbf{Modifying the concurrency algorithm} to optimize agent workloads based on Sab likelihood.
    \item \textbf{Adjusting prioritization} for suspected Sab customers in the service queue.
\end{enumerate}
These strategies rely on the company’s ability to identify suspected Sab customers in real time, using predictions based solely on customers' in-queue attributes---such as wait time, arrival time and day, customer details (e.g., returning customer), and the text of in-queue messages. A predictive model for this purpose can be developed following the approach outlined in Section \ref{sec:prob_abnd_WQ} (see Appendix \ref{app:ClassModQueue} for initial results of such a model).

The first strategy to reduce the Sab-related  capacity loss involves designing an AI agent (bot) to handle the suspected silent-abandonment conversations without requiring human agent involvement. This bot could manage the initial stages of the interaction, transferring the conversation to a human agent only if the customer responds. The bot could ask identifying questions or, with advancements in conversational AI, provide more sophisticated engagement tailored to the context of the service company. 
Even if the classification of suspected Sab customers is not perfect, implementing such a solution can help reduce wait times without changing staffing levels by decreasing the workload on human agents. 

The second strategy to mitigate Sab-related capacity loss is inspired by \cite{daw2021coproduction}, who suggested using within-conversation information to guide customer routing.  Building on this idea, we suggest modifying the concurrency algorithm to incorporate real-time conversation data, treating suspected Sab customers as fractional customers (rather than full ones) while they remain unresponsive. For example, a suspected Sab customer who has not written anything would count as half a customer. However, once the customer becomes responsive by sending a message, they would be updated to a full customer. Under this modified concurrency algorithm, an agent handling two suspected Sab customers and two fully responsive customers would be less busy than an agent managing four fully responsive customers and thus could be assigned an additional customer. This will reduce the amount of customers that are being delayed in the queue because of the Sab customers being ``served." 

A final possible solution we propose focuses on queue prioritization by suggesting the deprioritization of suspected Sab customers in the queue. Under this approach, when a suspected Sab customer's turn for service arrives, they would only be attended during the agent's low-concurrency period. This ensures that the impact of the Sab customers on overall system performance is minimized. This strategy is particularly advantageous for contact centers that do not operate 24/7. In such contact centers, customers may unknowingly join the queue when the contact center is closed, wait without receiving a response, lose patience, and silently abandon the queue. Since these abandonments are silent, when the center reopens the queue could be filled with Sab customers who consume agent capacity without actually being present. This scenario significantly delays service for new arrivals. By requiring suspected Sab customers to wait until an additional busy period has passed, the negative impact of Sab customers on system performance is mitigated. Although this prioritization policy may seem unfair towards the  suspected Sab  customers---who might still be waiting---it can be justified by the improvements in system efficiency and overall service quality.

\section{Estimating Customer Patience with Silent Abandonment}
\label{sec:patience_estimate}
So far, we have focused on identifying the Sab customers. However, Sab behavior also degrades data quality by creating missing information about abandonment times, rendering existing methods for estimating customer patience ineffective.
Our next goal is to develop estimators for customer patience in contact centers, addressing  the dual challenges of the uncertain classification of uSab customers to Sr1 or Sab and the missing  abandonment times for Sab customers. 
As mentioned in Section \ref{sec:introduction}, Sab introduces both left- and right-censoring into contact-center data on customer patience.  When the customer abandons the queue and provides an explicit indication of doing so---a known abandonment---they provide exact information regarding their patience and their patience corresponds to the observed wait time. However, the \emph{virtual wait time}, representing how long the customer would be required to wait had they  stayed in the queue, remains unknown. Therefore, patience serves as a lower bound for virtual wait time. 

When the customer is served, the wait time serves as a lower bound for their true patience. In this case, patience data is right-censored by the virtual wait time, itself uncensored. This type of right-censoring  was studied by \cite{Mandelbaum2013Data} using call-center data; we refer to their estimator as \emph{Method 1}.\footnote{Detailed formulas of Methods 1 and 2 are provided in Appendix \ref{app:mandelbaum_yefenot_formulas}.} In contrast to call centers, the contact centers' data regarding patience is also left-censored due to silent abandonment. When a customer silently abandons the queue---without providing an explicit indication---the observed wait time equals the virtual wait time (itself uncensored); thus, the wait time is an upper bound for the customer's true patience, which must be less than the wait time. \cite{Yefenof2018} addressed this type of left-censoring in the context of patients who left without being seen (LWBS) in emergency departments. We refer to their estimator as \emph{Method 2}.

However, due to incomplete data on customer classification, Method 2 cannot be directly applied to estimate patience in contact centers.  
Therefore,  we develop a new methodology for patience estimation that accommodates the complexity of missing data. 
Specifically, in Section \ref{sec:EM_model}, we develop an expectation-maximization (EM) algorithm tailored for estimating customer patience in contact centers. Section \ref{subsec:EM_validation} validates the algorithm's accuracy, sensitivity, and robustness. 
Later, in Section \ref{sec:EMextension-Model}, we extend the EM algorithm to a generalized model, enabling the estimation of how various factors, such as the number of words the customer writes while waiting, are associated with customer patience. 

\subsection{The EM Algorithm: Model Assumption and Formulation}
\label{sec:EM_model}
The problem of missing information on uSab customers stems from not knowing whether they received one-exchange service (Sr1), in which case their patience would be right-censored, or they silently abandoned, in which case their patience would be left-censored. Nonetheless, we know that the time these customers waited in the queue, and hence their virtual wait time, is uncensored. 

Following the formulation of \cite{Yefenof2018}, let $\tau$ be customer-patience time (failure time) and assume it has a cumulative distribution function (CDF) $F$ and a probability distribution function (PDF) $f$. Assume that $\tau\sim \exp(\theta)$. This assumption follows  \cite{Brown2005}, who showed using call-center data (with no delay announcements) that patience distribution has an exponential tail. 
Let $W$ be the virtual wait time (censoring time)---the time the customer is required to wait by the system---and assume it has a CDF $G$ and a PDF $g$.
We know from queueing theory that in overloaded systems, like the contact centers we are investigating, wait time is close to exponentially distributed \citep{Kingman1962}. In addition, \cite{Brown2005} showed that in call centers with no delay announcements, virtual wait time is close to exponentially distributed. In our dataset, we also have no delay announcements that may change customer patience while waiting \citep{Ibrahim2018DelayAnnReview}; hence, we can make a realistic assumption that the virtual waiting times are exponentially distributed. 
Formally, assume that $W\sim \exp(\gamma)$.
Let $\Delta$  be an indicator for the case where the customer loses patience before the agent replies: $\Delta \triangleq 1_{\{\tau\leq W\}}$. Conversations in which information regarding $\Delta_i$ is missing are assigned a null value. 
Let $Y$ be a random variable indicating whether the customer will inform the system when abandoning. We assume that $Y\sim \mathrm{Bernoulli}(q)$, where $q$ is the probability that the customer will inform the system when abandoning; formally, $q \triangleq \textit{Pr}\left\{ \textit{Indicate\;abandonment}\right\}$.

Assume that $W$ and $\tau$ are independent, as is frequently done in right-censoring survival analysis (e.g., \citealp{smith_2002,Mandelbaum2013Data,Yefenof2018}). Moreover, this is a natural assumption in contact centers since patience is decided by the individual customer while the virtual wait time is decided by the company. This is indeed the case in our contact centers where no delay information, such as their place in queue, is provided to the customer.
Additionally, we assume that $Y$ and $W$ are independent. That is, the decision of a customer to indicate whether they abandoned the queue is independent of their wait time. For example, customers might tend to leave windows open in the computer even if they are not using them; therefore, this tendency would be independent of the wait time. The independence assumption of $Y$ and $W$ is for tractability reasons; currently, we do not have evidence to support this assumption and suggest that it be relaxed in future research. Finally, let $U$ be the system's observed time. For each arriving customer $i$, we observe the vector of data $(U_{i},Y_{i},\Delta_{i})$, $i=1,...,n$. We also assume that our sample of customers is homogeneous: all customers have the same patience parameter $\theta$ and abandonment indicator $q$.

Summarizing, our model rests on the following assumption:
\begin{assumption} \label{assumption:1}
The EM algorithm developed in Section \ref{sec:EM_model} assumes
(1) Patience time is $\tau\sim \exp(\theta)$,
(2) Virtual wait time is $W\sim \exp(\gamma)$,
(3) Customer abandonment indicator is $Y\sim \mathrm{Bernoulli}(q)$,
(4) $W$ and $\tau$ are independent,
(5) $Y$ and $W$ are independent, and
(6) Customers are homogeneous.
\end{assumption}

\subsubsection{Customer Classes with Complete Data.}
\label{subsec:ClassComplete}
In Table \ref{tbl:CompeteDtaNotation}, we follow \cite{Yefenof2018} in formally defining three customer classes under the assumption of complete data regarding which customers abandoned. The table identifies each customer class by type, notation indicator, formal definition of that indicator (based on values $\Delta$ and $Y$), what variable is observed in $U$, and what variable is censored by $U$ and in what direction. 

\begin{table}[!htb]
  \centering
  \caption{Classes of Customers: Complete Data}
  \begin{small}
  \begin{tabular}{lcccccl} 
  \toprule
  Class Type & Notation Indicator & Formal Definition & $\Delta$ & $Y$ & $U$ Observed & Censored (direction)\tabularnewline
  \midrule 
  Service &  $C_{1}=1$ & $1-\Delta\equiv 1$ & 0 & 0 & $W$ & $\tau$ (right-censored)\tabularnewline
  Kab & $C_{2}=1$ & $Y\Delta$ & 1 & 1 & $\tau$ & $W$ (right-censored) \tabularnewline
  Sab & $C_{3}=1$ & $(1-Y)\Delta$ & 1 & 0 & $W$ &$\tau$ (left-censored) \tabularnewline
  \bottomrule

  \end{tabular}
  \end{small}

  \label{tbl:CompeteDtaNotation}
\end{table}


\subsubsection{Customer Classes with Missing Data.} \label{subsec:ClassesMissingDa}
Due to the problem of missing data on the uSab conversations, we are not able to categorize all the conversations into just one of the classes we defined in Section \ref{subsec:ClassComplete}. Therefore, we need to formulate additional class indicators. Let $M$ denote the customer classes in a system in which there is missing data on which individual customers abandoned. These classes are defined in Table \ref{tbl:MissingDtaNotation}.

\begin{table}[!htb]
  \centering
  \caption{Classes of Customers: Missing Data}
  \begin{small}
  \begin{tabular}{lcccccl} 
  \toprule
  Class Type & Notation Indicators & Formal Definition & $\Delta$ & $Y$ & $U$ Observed & Censored (direction) \tabularnewline
  \midrule 
  Service & $C_1=1; M=1$ & $1-\Delta\equiv 1$ & 0 & 0 & $W$ & $\tau$ (right-censored)\tabularnewline
  Kab & $C_2=1; M=2$ & $Y\Delta$ & 1 & 1 & $\tau$  & $W$ (right-censored) \tabularnewline
  uSab: &  &  &  &  &  &  \tabularnewline
  ~~~uSab is Sab & $C_3=1;M=0$ & $(1-Y)\Delta$ & null & 0 & $W$  & $\tau$ (left-censored) \tabularnewline
  ~~~uSab is Sr1 & $C_1=1;M=0$ & $1-\Delta\equiv 1$ & null & 0 & $W$  & $\tau$ (right-censored) \tabularnewline
  \bottomrule

  \end{tabular}
  \end{small}

  \label{tbl:MissingDtaNotation}
\end{table}
%

\subsubsection{The EM Algorithm Formulation.}
\label{subsec:EM_algorithm}
The EM algorithm estimates the following parameters simultaneously: the rate at which customers lose patience, $\theta$; the probability of informing the system when abandoning, $q$; and the rate of the virtual wait time distribution, $\gamma$. The optimization problem is defined to maximize the likelihood function, which measures the probability that the observations are given from the assumed distributions given the parameters ($\theta, q, \gamma$). We write the likelihood of the observed data $D\triangleq \{(U_{i},Y_{i},\Delta_{i}),$  $i=1,...,n\}$ as follows:
\begin{equation}
\begin{aligned} \label{eq:Likelihood}
L(D;\theta,q,\gamma)= & \prod_{i=1}^{n}\left\{ e^{-\theta U_{i}}\gamma e^{-\gamma U_{i}}\right\} ^{C_{1}^{i}}\left\{ q\theta e^{-\theta U_{i}}e^{-\gamma U_{i}}\right\} ^{C_{2}^{i}}\left\{ (1-q)(1-e^{-\theta U_{i}})\gamma e^{-\gamma U_{i}}\right\} ^{C_{3}^{i}}\\
= & \prod_{i=1}^{n}\left\{ e^{-\theta U_{i}}\gamma e^{-\gamma U_{i}}\right\} ^{1-\Delta_{i}}\left\{ q\theta e^{-\theta U_{i}}e^{-\gamma U_{i}}\right\} ^{\Delta_{i}Y_{i}}\left\{ (1-q)(1-e^{-\theta U_{i}})\gamma e^{-\gamma U_{i}}\right\} ^{(1-Y_{i})\Delta_{i}.}
\end{aligned}
\end{equation}
The function is formulated following \cite{Yefenof2018}: the first part is for the served customer ($C_{1}^{i}=1$), where we multiply the survival function of the customer patience ($1-F_{\tau}\left(u\right)$) by the PDF of the customer's wait time. The second part is for the Kab customer ($C_{2}^{i}=1$), where we multiply the probability of informing when abandoning by the PDF of the customer patience and the survival function of the customer's wait time ($1-G_{W}\left(u\right)$). Finally, the third part is for the Sab customer ($C_{3}^{i}=1$), where we multiply the probability of not informing when abandoning by the CDF of the customer patience and the PDF of the customer's wait time.  

However, this likelihood function depends on knowing the complete data. Recall that some of the observations belong to the class $M=0$ since they have missing data in $\Delta$. Therefore, we cannot find the parameters by simply solving the maximization problem. 
Instead, we need to formulate an EM algorithm (see Algorithm \ref{EM}), a well-known computing strategy for dealing with problems of missing data and censoring
\citep{little2002statistical}. The algorithm estimates the parameters ($\theta, q, \gamma$) 
using Theorems \ref{Theorem1} and \ref{Theorem2}. Specifically, it estimates starting parameter values and subsequently iterates between the expectation step (E-step)---using Theorem \ref{Theorem1}---and the maximization step (M-step)---using Theorem \ref{Theorem2}---and updates these estimators until convergence.  
In the $t$th iteration, the E-step consists of finding a surrogate function (given in Equation~\eqref{eq:loglikelihoodMisg}) that is a lower bound on the log-likelihood function (given in Equation~\eqref{eq:loglikelihood}) and is tangent to the log-likelihood at $(\widehat{\theta^{(t)}},\widehat{q^{(t)}},\widehat{\gamma^{(t)}})$. In practice, it is enough to compute the expectation of the log-likelihood given the information of the previous iteration, which is presented in Equation \eqref{eq:ci} of Theorem \ref{Theorem1}.

\begin{equation}
\begin{aligned}\label{eq:loglikelihoodMisg}
l(D,\theta,q,\gamma) & =\sum_{i=1}^{n}\left\{ \left(\widehat{C_{1,t}^{i}}\right)\left(\log\gamma-\gamma U_{i}-\theta U_{i}\right)\right\} \\
 & +\sum_{i=1}^{n}\left\{ \left(\widehat{C_{2,t}^{i}}\right)\left(\log\theta-\theta U_{i}-\gamma U_{i}+\log q\right)\right\} \\
 & +\sum_{i=1}^{n}\left\{ \left(\widehat{C_{3,t}^{i}})\right)\left[\log\left(1-q\right)+\log(1-e^{-\theta U_{i}})+\log\gamma-\gamma U_{i}\right]\right\} .
\end{aligned}
\end{equation}

\begin{algorithm}
\SetAlgoLined
\KwResult{$\widehat{\theta^{(t+1)}}$, $\widehat{q^{(t+1)}}$, and $\widehat{\gamma^{(t+1)}}$.}

\vspace{3pt}
Initialization: For every customer $i$, use Equation \eqref{eq:ci} to calculate $\widehat{C_{1,0}^{i}}$ and $\widehat{C_{2,0}^{i}}$ and $\widehat{C_{3,0}^{i}}=\hat{\pi}_{i} 1_{\{M^{i}=0\}}$, where $\hat{\pi_{i}}\in\left[0,1\right]$ is chosen randomly. 
To obtain the starting
parameters, $(\widehat{\theta^{(1)}},\widehat{q^{(1)}},\widehat{\gamma^{(1)}})$, solve Equations \eqref{eq:PartialDTheta} and \eqref{eq:PartialDGammaq}, respectively. Set $t=0$.\\
\vspace{6pt}
 \While{ $ \mid\widehat{\theta^{(t)}}-\widehat{\theta^{(t+1)}}\mid+\mid\widehat{q^{(t)}}-\widehat{q^{(t+1)}}\mid+\mid\widehat{\gamma^{(t)}}-\widehat{\gamma^{(t+1)}}\mid>\epsilon$}{
  E-step: Given the observed data
  $D=\{(U_{i},Y_{i},\Delta_{i})$ $i=1,...,n\}$ and the current estimations of the parameters $(\widehat{\theta^{(t)}},\widehat{q^{(t)}},\widehat{\gamma^{(t)}})$, compute $\widehat{C_{j,t}^{i}}$, $j=1,2,3$ $\forall i=1,...,n$ using Eq.\ \eqref{eq:ci}. \\
\vspace{3pt}
  M-step: Maximize to obtain $(\widehat{\theta^{(t+1)}},\widehat{q^{(t+1)}},\widehat{\gamma^{(t+1)}})$.\
  That is, update the estimations of the parameters using Equations \eqref{eq:PartialDTheta} and \eqref{eq:PartialDGammaq}, respectively.\\
  \vspace{3pt}
  Update $t\leftarrow t+1$.
 }
 \caption{The EM Algorithm}
 \label{EM}
\end{algorithm}


\begin{theorem} \label{Theorem1}
Under Assumption \ref{assumption:1}, $\widehat{C_{1,t}^{i}}$, $\widehat{C_{2,t}^{i}}$, and $\widehat{C_{3,t}^{i}}$ are given by
\begin{equation} 
\begin{split}
\label{eq:ci}
&\widehat{{C_{1,t}^{i}}}=(1-\widehat{C_{3,j}^{i}})1_{\{M^{i}=0\}}+1_{\{M^{i}=1\}};  \\
&\widehat{C_{2,t}^{i}}=1_{\{M^{i}=2\}}; \\
&\widehat{C_{3,t}^{i}}=1_{\{M^{i}=0\}}\left(1-e^{-\widehat{\theta^{(t)}}U_{i}}\right).
\end{split}
\end{equation}
\end{theorem}
The proof is given in Appendix \ref{app:EMproofs-ExpectationStep}.

The notations $\widehat{C_{1,t}^{i}}$, $\widehat{C_{2,t}^{i}}$, and $\widehat{C_{3,t}^{i}}$ represent the probabilities (weights) for the $i$th customer to belong to class $C_1, C_2,$ or $C_3$, respectively, given the parameters from iteration $t-1$, $(\widehat{\theta^{(t)}},\widehat{q^{(t)}},\widehat{\gamma^{(t)}})$, and the observed data. Note that the EM algorithm's update of the weights with missing data in the $t-1$ iteration, $\widehat{C_{j,t-1}^{i}}$
$j=1,3$, is different for each observation $i$ in the data class $M^{i}=0$. 
That is, $\widehat{C_{3,t-1}^{i}}$ need not equal $\widehat{C_{3,t-1}^{k}}$, given that $M^{i}=M^{k}=0.$
In the M-step of the $t$th iteration, $(\widehat{\theta^{(t+1)}},\widehat{q^{(t+1)}},\widehat{\gamma^{(t+1)}})$ are found (in Equations \eqref{eq:PartialDTheta} and \eqref{eq:PartialDGammaq}, respectively) to be the maximizers of the surrogate function in Equation \eqref{eq:loglikelihoodMisg}.

\begin{theorem} \label{Theorem2}
Under Assumption \ref{assumption:1}, the parameters $\widehat{q^{(t+1)}}$,  $\widehat{\gamma^{(t+1)}}$ are given by
\begin{equation}
\begin{split} \label{eq:PartialDGammaq}
&\widehat{q^{(t+1)}}=\left\{ \sum_{i=1}^{n}\widehat{C_{2,t}^{i}}\right\} \left\{ \sum_{i=1}^{n}\left(1-\widehat{C_{1,t}^{i}}\right)\right\} ^{-1},  \\
&\widehat{\gamma^{(t+1)}}=\left\{ \sum_{i=1}^{n}\left(1-\widehat{C_{2,t}^{i}}\right)\right\} \left\{ \sum_{i=1}^{n}U_{i}\right\} ^{-1}, 
\end{split}
\end{equation}
and the parameter $\widehat{\theta^{(t+1)}}$ is given as a solution to the following equation:
\begin{equation}\label{eq:PartialDTheta}
\widehat{\theta^{(t+1)}}\left\{ \sum_{i=1}^{n}\left(\widehat{C_{3,t}^{i}}-1\right)U_{i}\right\} +\sum_{i=1}^{n}\widehat{C_{2,t}^{i}}+\widehat{\theta^{(t+1)}}\left\{ \sum_{i=1}^{n}\widehat{C_{3,t}^{i}}\frac{U_{i}e^{-\widehat{\theta^{(t+1)}}U_{i}}}{1-e^{-\widehat{\theta^{(t+1)}}U_{i}}}\right\} =0.
\end{equation}

\end{theorem}
The proof is given in Appendix \ref{app:EMproofs-MaximizationStep}.

We repeat the E-step and the M-step until convergence for some predetermined $\epsilon>0$. 
The procedure ends when we find a maximum of  the likelihood function that yields estimators for the parameters $(\widehat{\theta^{(t+1)}},\widehat{q^{(t+1)}},\widehat{\gamma^{(t+1)}})$.

Regarding convergence of our EM algorithm, by Theorem 8.1 of \cite{little2002statistical}, each iteration of the EM algorithm increases the likelihood function. Then, because the smoothness conditions hold for the exponential distribution, the sequence of estimators converges to a stationary point by Theorem 8.2.

%
More details on the EM algorithm and proofs are provided in Appendix \ref{app:EMproofs}.


\subsection{Validation of the EM Algorithm}
\label{subsec:EM_validation}
We perform several performance evaluations to validate the use of our EM algorithm in practice. Due to space constraints, most of the validation tests are described  in Appendix \ref{app:validation}. In Appendix \ref{app:EMaccuracy}, we compare the accuracy of the EM algorithm to previous methods of estimating customer patience, \cite{Mandelbaum2013Data} (Method 1) and \cite{Yefenof2018} (Method 2). These basic comparisons use simulated data and demonstrate that our EM algorithm provides the most accurate estimation of all parameters ($\theta, \gamma$, and $q$), regardless of the level of load in the system. 
Here, in Section \ref{subsec:EMValidation_data}, we validate the accuracy of the EM estimators using \emph{real data}, concluding that the EM algorithm is the only method that provides an accurate estimation of customer patience in reality. 
In Appendix \ref{app:EMSensitivity}, we examine the algorithm's sensitivity under the initial conditions. We show that the parameter estimations are stable and do not change when different initial values are inserted in the EM algorithm. This suggests that one does not need to use the output of the classification model we developed in Section \ref{sec:prob_abnd_WQ}  (or any model with similar sensitivity and specificity proportions) as starting probabilities in the EM algorithm. 
(In all tests throughout this paper, we set $\epsilon =10^{-6}$.)

As mentioned, the EM algorithm can cope with the missing data, but Methods 1 and 2 cannot.
To use them as benchmark methods for these comparisons, we must make certain assumptions on how they cope with uSab conversations ($M=0$). To apply  \cite{Yefenof2018}, we have two options of how to classify uSab conversations: either as served (Sr1) customers ($C_{1}=1$) or as Sab customers ($C_{3}=1$). 
To apply the method of \cite{Mandelbaum2013Data}, we can classify uSab conversations as either served customers ($C_{1}=1$) or Kab ($C_{2}=1$), since this method cannot deal with left-censored conversations. These options result in all possible (four) baseline methods used for our accuracy comparisons.

\subsubsection{Estimating Patience in the XYZ Dataset: Accuracy and Robustness Tests.}
\label{subsec:EMValidation_data}
Using the XYZ contact-center dataset described in Section \ref{sec:dataAndReseatch}, we compare the EM algorithm to four benchmark methods for estimating customer patience.
The results are presented in Table  \ref{tbl:AvPatience}. 

\begin{table}[!htb]
  \centering
  \caption{Comparison of Estimations of Average Customer Patience: Telecommunications Dataset (May 2017)}
  \begin{scriptsize}
  \begin{tabular}{llcc} 
  \toprule
  Row & Method & Avg.\ Patience (Minutes) & Data used\tabularnewline
  \midrule
  1 & Method 2---Using sample of labeled conversations & ~81.90 & Sample$^*$\tabularnewline
  2 & Method 2---With SVM classification$^{**}$  & ~69.32 & Full data\tabularnewline
  3 & Method 1---Uncertain silent abandonment is service & 166.42 & Full data\tabularnewline
  4 & Method 2---Uncertain silent abandonment is service & 188.07& Full data\tabularnewline
  5 & Method 1---Uncertain silent abandonment is abandonment  & ~28.27& Full data\tabularnewline
  6 & Method 2---Uncertain silent abandonment is silent abandonment & ~13.17& Full data\tabularnewline
  7 & {EM} & {~81.11}& Full data\tabularnewline
\bottomrule
\multicolumn{4}{l}{$^*$The sample includes 2500 conversations, including 650 uSab conversations that were manually labeled to Sab or Sr1 classes.}\\
\multicolumn{4}{l}{$^{**}$uSab conversations classified with SVM cutoff of 0.47.}
  \end{tabular}
  \end{scriptsize}

  \label{tbl:AvPatience}
\end{table}


The main challenge in comparing the methods is the lack of ground truth for customer patience. To overcome this challenge, we utilized the manually tagged data described in Section  \ref{sec:prob_abnd_WQ}.  
Since this labeled dataset contains complete information on which customers abandoned the queue, we could apply the method from \cite{Yefenof2018} to estimate customer patience. The resulting  patience estimate was 81.9 minutes (row 1 of Table \ref{tbl:AvPatience}), serving as a reference point for comparison. 

The EM algorithm's patience estimate, based on the full monthly dataset, closely matched the estimate derived from the labeled data at 81.11 minutes (row 7 of Table \ref{tbl:AvPatience}). In contrast, the other benchmark methods (rows 3--6) produce significantly divergent estimates (13--188 minutes). Implementing Method 2 on the classified data (using the SVM classifier with a threshold of 0.47) provides better results with an estimated patience of 69.32 minutes (row 2 of Table \ref{tbl:AvPatience}), yet even this result is biased, probably due to some conversation misclassification. This result demonstrates that the EM algorithm (Algorithm \ref{EM}) is the only one that effectively handles the complexities of missing data, providing an accurate estimate of  patience. 
 
A closer look at Table \ref{tbl:AvPatience} highlights the substantial bias introduced by missing data. When silent abandonment and missing data are ignored---regarding all uSab cases ($M=0$) as serviced ($C_{1}$=1)---and  customer patience is estimated using either Method 1 or Method 2,  patience is overestimated by twice or more (rows 3 and 4 of Table \ref{tbl:AvPatience}). This reflects the current practice of many companies, which often rely on Method 1 (row 3) while ignoring the concept of silent abandonment that creates left-censoring and missing data.

More advanced companies might recognize the phenomenon of silent abandonment but fail to account for missing data. These companies may classify all conversations in class $M=0$ as Sab ($C_{3}=1$) and then apply Method 1 (ignoring left-censoring) or Method 2 (accounting for left-censoring). Both approaches lead to an underestimation of customer willingness to wait (rows 5 and 6).

Our finding that customers in the XYZ contact center exhibit a willingness to wait for over an hour (row 1 of Table \ref{tbl:AvPatience}) may seem surprising, but it aligns with the following observations:
\begin{enumerate}
    \item[(1)] \textbf{Customer expectations:} The XYZ contact center sends an automatic message advising customers to ``go on with their daily activities" (while waiting for a reply) and treat the interaction as if ``talking to a friend." This sets expectations for longer wait times and adjusts customer patience accordingly.  
    \item[(2)] \textbf{Customer familiarity:} XYZ Company has a high proportion of returning customers due to an ongoing relationship. These customers have realistic expectations of the virtual wait time, which averages 8.77 minutes. The fact that customer patience exceeds the virtual wait time is consistent with findings from the call-centers literature \citep{Brown2005}. 
    Furthermore, returning customers may be aware of the company's automatic closure policy. As stated in Section \ref{sec:dataAndReseatch}, the contact center automatically closes communication after two hours of inactivity. A returning customer familiar with this policy may feel less pressured to stay alert,  reducing anxiety and stress---factors known to decrease patience \citep{Maister2005TheLines}.
    \item[(3)] \textbf{Service requirement vs.\ patience:} \cite{Mandelbaum2013Data} showed that customers are often willing to wait around two (or more) times longer than their service requirement. With a service time of 46.34 minutes in this contact center, the estimated patience aligns with this ratio. 
    \item[(4)] \textbf{Writing while waiting:} The ability to send messages while waiting positively influences patience, as it creates a sense of busyness and engagement  \citep{Maister2005TheLines}. We explore this relationship in Section \ref{sec:patience_coef}. 
\end{enumerate}

\section{The Influence of System Design on Customer Behavior and Silent Abandonment 
}
\label{sec:patience_coef}
Referring back to Figure \ref{fig:cust_Process_flow_WQ}, we note that the ambiguity surrounding uSab customers could be resolved by restricting the ability to write while waiting. Under such a policy, the Sr1 customer class would no longer exist, as single-interaction services require customers to communicate their issues explicitly. However, the Sab customer class remains well-defined: customers who abandon the system without a clear indication of doing so would still be categorized as Sab (even when not writing their inquiry).

Indeed, contact-center infrastructure allows for such restrictions, and some companies opt to implement them. This approach simplifies the analysis of performance measures, eliminates the need for sophisticated classification models, and enables identification of Sab customers based solely on metadata.

However, this raises critical questions: \emph{What are the benefits of allowing customers to write while waiting? Should companies adopt such restrictions?}

We argue that while prohibiting customers from writing during the wait can reduce some of the challenges associated with missing data (e.g., conversation misclassification), this approach would incur significant costs. Specifically, it could harm the customer experience, reduce patience, and increase abandonment. 

To explore these trade-offs, we first present the rationale behind these claims. Then, in Section \ref{sec:EMextension-Model}, we extend the EM algorithm to model how different variables influence customer patience. Subsequently, in Section \ref{sec:EMextention-Patience Estimation}, we apply this extended model to quantify the impact of allowing customers to write while waiting. Finally, in Section \ref{sec:BoostingExpiriance}, we propose managerial strategies to enhance customer satisfaction and minimize abandonment based on these findings.

From a broader perspective, allowing customers to write while waiting enhances their engagement and experience within the system. This policy facilitates a communication environment similar to instant messaging platforms (e.g., WhatsApp), where there are no constraints on the timing of messages. To support the claim that people prefer to engage with the system while waiting, we measured customer writing behavior in the XYZ dataset. On average, customers send 1.2 messages while waiting ($SD = 0.51$), which accounts for 20\% of their total messages and 
 27.8\% of their total written words (see data in Table \ref{tbl:general_stat}).

This ability to write while waiting has operational benefits, particularly in reducing agent service time. When customers use their waiting time to provide information, agents have most of the relevant information needed to solve the customer problem once assigned. This reduces the time spent on each interaction.
Currently, the average service time for Sr customers in XYZ Company is 53.8 minutes, with 1.2 out of 7.9 total messages (15\%) written while waiting in the queue. If queue writing was disallowed, these messages would shift to the service phase, increasing the service time by 13.2\%, from 53.8 to 62.0 minutes.

We hypothesize that the ability to write while waiting also increases customer patience. This idea draws on the behavioral operations literature, specifically \citet{Maister2005TheLines}, who argues that when people are occupied, they are less likely to perceive waiting as burdensome,  resulting in greater patience. A good way to occupy customer time is to enable them to start the service while waiting by explaining their problem to the company. Therefore, allowing customers to write while waiting can be viewed as a way to transfer customer perception of wait time to service time, which reduces perceived waiting and increases patience. Additionally, a sunk cost effect may apply. \cite{Ulku2022SocialQueues} found that customers who invest more time in the queue are less likely to abandon the system. Similarly, we can conjecture that customers who invest more effort in writing their inquiry while waiting are less likely to abandon the queue (silently).  
To test this hypothesis, we extend the EM algorithm to include covariates that capture the impact of writing behavior on customer patience.



\subsection{The Extended EM Algorithm} \label{sec:EMextension-Model}
Let the patience parameter $\theta$ be influenced by $k$ variables denoted by $\textbf{X}=[X_{1},...,X_{k}]$, such that  
\begin{equation}
    \theta | \textbf{X}\triangleq e^{-(\beta_{0}+\beta_{1}X_{1}+...+\beta_{k}X_{k})}=e^{-(\beta_{0}+\beta^{T} \textbf{X})}.
\end{equation} 
This expression represents an exponential function of a linear combination of the coefficients $\beta_{1},...,\beta_{k}$  and $k$ variables, forming a structure known as a log-linear or exponential regression model, commonly used in survival analysis \citep{kleinbaum2012survival}. In this model, $\beta_0$ serves as the intercept term, representing the baseline level of $\theta$ in the absence of contributing variables. The other terms in the exponential function adjust that baseline level with the magnitude and direction determined by the sign and value of the respective $\beta_j$. This log-linear model implies that each unit increase in a predictor $X_j$ has a multiplicative effect on the parameter $\theta$. 
Specifically, a one-unit increase in $X_j$ scales the mean patience time by a factor of $e^{-\beta_j}$. The effect on customer patience (the inverse of $\theta$) is as follows:
\begin{itemize}
    \item If $\beta_j>0$, an increase (or decrease) in $X_j$ will increase (or decrease) customer patience.
    \item If $\beta_j<0$, an increase (or decrease) in $X_j$ will decrease (or increase) customer patience.
\end{itemize}
Using notation similar to Section \ref{subsec:EM_algorithm}, the observed data is now represented as $D \triangleq \{(U_{i},Y_{i},\triangle_{i},X_{i,1},...,X_{i,k})$, $i=1,...,n\}$. With this setup, and after simplification, the log-likelihood function 
becomes (for details, see Appendix \ref{app:EMextantion-proofs})

\begin{equation} \label{eq:log_lik_cov}
\begin{aligned}l(D,\beta,q,\gamma) & =\sum_{i=1}^{n}\left\{ \left(1-\Delta_i\right)\left(-U_{i}e^{-(\beta_{0}+\beta^T\textbf{X}_{i})} + \log\gamma \right)\right\} +\sum_{i=1}^{n}\left\{ -\gamma U_{i}\right\} \\
 & +\sum_{i=1}^{n}\left\{ \left(\Delta_{i}Y_{i}\right)\left[-\beta_{0}-\beta^T\textbf{X}_{i}- U_{i}e^{-(\beta_{0}+\beta^T\textbf{X}_{i})}+\log(q)\right]\right\} \\
 & +\sum_{i=1}^{n}\left\{ \left((1-Y_{i})\Delta_{i}\right)\left[\log(1-q)+\log\left(1-e^{-U_{i}e^{-(\beta_{0}+\beta^T\textbf{X}_{i})}}\right)\right]\right\},
\end{aligned}
\end{equation}
where
 $\beta\triangleq\begin{bmatrix}\beta_{1} & \cdots & \beta_{k}\end{bmatrix}$ and 
$\textbf{X}_{i}\triangleq\begin{bmatrix}X_{i,1} & \cdots & X_{i,k}\end{bmatrix}$.

To maximize the likelihood with missing data on $\Delta$, we extended the EM algorithm (see Algorithm \ref{EMCV}). 
In the expectation step, we  calculate $\widehat{C_{1,t}^{i}}$, $\widehat{C_{2,t}^{i}}$, and $\widehat{C_{3,t}^{i}}$ using Equation \eqref{eq:ci} as in Algorithm \ref{EM} but now with $\theta | \textbf{X} = e^{-(\beta_{0}+\beta^{T} \textbf{X})}$. In the maximization step, the extended EM algorithm  estimates $\gamma$, $q$, and the coefficients $\beta_{0}$ through $\beta_{k}$ using Equations \eqref{eq:PartialDGammaq} and \eqref{eq:Betas}, respectively. We verified the extended EM algorithm using simulated data similar to that described in Section \ref{subsec:EM_validation}. The validation checks were successful (details omitted for brevity). Further information about the EM algorithm extension is provided in Appendix \ref{app:EMextantion-proofs}.

\subsection{Estimating the Associations of Patience with System Design and Customer Behavior} \label{sec:EMextention-Patience Estimation}
We implement the extended EM model to the XYZ database in order to investigate the association between customer patience and their in-queue behavior. 
Our main variable of interest is the number of words the customer types while waiting in the queue. To address the variations in word counts and reduce the influence of extreme cases, we categorized this variable in two ways: 
\begin{enumerate}
    \item \textbf{Binary variable:} Named \emph{\# Words in Queue: Binary}, this variable is defined as ``0'' if the customer wrote up to one word while waiting and ``1'' otherwise (i.e.,\ for two or more words).
\item  \textbf{Categorical variable:} Named \emph{\# Words in Queue: Categorical}, this variable has seven levels: $\leq 1$ (baseline), 2--10, 11--20, 21--30, 31--40, 41--50, and 50+ words. 
\end{enumerate}
Writing up to one word happens either when the customer chooses not to engage with the agent during their wait or when the text is non-informative (e.g., ``hello''). 

We included several control variables to account for external factors that might influence customer patience:
\emph{Communication Platform}, whether the customer used the company’s app (``0'', baseline) or the website (``1''); \emph{Returning Customer}, whether the customer is a returning user (``0'' is No, baseline; ``1'' is Yes); \emph{Day of the Week}, a categorical variable with the weekend (Saturday and Sunday combined) as the baseline; \emph{Time of Day}, divided into four 6-hour segments to capture hourly variations in behavior; \emph{Customer Sentiment}, expressed sentiment of the customer, categorized as neutral (baseline), negative, or positive.
Customer sentiment was included since it influences contact-center efficiency through agent and customer behavior \citep{Altman2021} and reflects customer satisfaction \citep{Ashtar2024Affect-as-Information}. This variable was measured using the CustSent tool  \citep{YomTov2018CustSent}.
%
%
It is important to note that the variables included in the models are restricted to information available while the customer is in the queue. This ensures that the analysis captures effects relevant to the customer's real-time experience in the system, emulating a scenario where the company can make informed decisions during the wait. 

We implemented two separate extended EM models corresponding to the two definitions of the ``\# words in queue'' variable. The results of these models are summarized in Table \ref{tbl:coef}.
For each variable, we report its coefficient ($\beta$) and calculate confidence intervals  using 500 bootstrapped samples.  The lower and upper bounds of these intervals are also provided in Table \ref{tbl:coef}.
A variable is considered statistically significant if its confidence interval does not include zero, and such variables are marked with an asterisk ($^*$) next to the coefficient.
Additionally, we quantify the change in patience level when a variable’s category is active, offering practical insights into how customer attributes and behaviors influence patience in the queue.

\begin{table}[!htb]
\centering
 \caption{Results of EM Extension for Estimating Customer Patience. XYZ Dataset (May 2017)} 
\begin{scriptsize}
\begin{tabular}{llcccccccc}
\toprule
\multicolumn{2}{c} {Variable} && \multicolumn{3}{c}{Binary} & & \multicolumn{3}{c}{Multiple}  \\
\cmidrule{1-2} \cmidrule{4-6} \cmidrule{8-10} 
Name & Category  && Coef.\ & (Lower, Upper) & Patience chg.\ && Coef.\ & (Lower, Upper) & Patience chg.\ \\ \midrule
Intercept&   && $~3.438^*$  & (~2.979, ~3.896) & 31.12 && $~3.613^*$  & (~3.333, ~3.892)   & 37.06\\
\multicolumn{2}{l}{\# Words in Queue:} && & & && & & \\
Binary & 2+ && $~1.052^*$ & (~0.944, ~1.160) & 2.86 &&  &  & \\
Categorical & 2--10 && & & && $~0.756^*$  & (~0.626, ~0.886) & 2.13 \\
 & 11--20 && & & && $~1.120^*$  & (~0.991, ~1.250) & 3.07 \\
 & 21--30 && & & && $~1.357^*$  & (~1.207, ~1.508) & 3.89 \\
 & 31--40 && & & && $~1.537^*$  & (~1.329, ~1.746) & 4.65 \\
 & 41--50 && & & && $~1.654^*$  & (~1.402, ~1.906) & 5.23 \\
 & 51+ && & && & $~1.829^*$  & (~1.612, ~2.046) & 6.23 \\ 
\multicolumn{2}{l}{\emph{Control variables:}} && & & && & & \\
Communication Platform & Web  && -$0.138^*$ & (-$0.241$, -$0.035$) & 0.87 && -$0.240^*$ & (-$0.344$, -$0.135$)  & 0.79 \\
Returning Customer & Yes  && $~0.891^*$  & (~0.809, ~0.973)   & 2.44 && $~0.862^*$  & (~0.763, ~0.960)   & 2.37 \\
Day of Week & Mon  && $~0.213^*$  & (~0.065, ~0.361)   & 1.24 && $~0.203^*$  & ($~0.042$, ~0.364)   & 1.22 \\
 & Tue  && $~0.494^*$  & (~0.377, ~0.611)   & 1.64 && $~0.457^*$  & (~0.325, ~0.589)   & 1.58 \\
 & Wed  && $~0.253^*$  & (~0.110, ~0.396)   & 1.29 && $~0.230^*$  & (~0.074, ~0.386)   & 1.26 \\
 & Thu && 0.064 & (-$0.080$, ~0.209)   & 1.07 && 0.065 & (-$0.114$, ~0.243)   & 1.07 \\
 & Fri  && -$0.266^*$ & (-$0.456$, -$0.077$) & 0.77 && -$0.261$~ & (-$0.400$, -$0.121$) & 0.77 \\
Time of Day & 01:00--07:00 && 0.188 & (-$0.058$, ~0.434)   & 1.21 && 0.155 & (-$0.054$, ~0.363)   & 1.17 \\
 & 07:00--13:00 && $~0.396^*$  & (~0.254, ~0.537)   & 1.49 && $~0.370^*$  & (~0.241, ~0.498)   & 1.45 \\
 & 13:00--19:00 && $~0.226^*$  & (~0.116, ~0.336)   & 1.25 && $~0.204^*$  & (~0.107, ~0.302)   & 1.23 \\
Customer Sentiment& Negative && -$0.473^*$ & (-$0.690$, -$0.257$) & 0.62 && -$0.558^*$ & (-$0.786$, -$0.331$) & 0.57 \\
 & Positive && $~0.259^*$  & (~0.132, ~0.386)   & 1.30 && -0.008~ & (-0.131, ~0.114)  & 0.99  \\ \bottomrule
\multicolumn{9}{l}{$^*$ Confidence interval does not contain 0.}
\end{tabular}
\label{tbl:coef}
  \end{scriptsize}
\end{table}

Both models in Table~\ref{tbl:coef} consistently demonstrate a strong association between writing more than one word while waiting in queue and increased customer patience. In the binary word categorization model, customers who type more than one word exhibit a potential  2.86-fold increase in patience. Similarly, in the multiple-level categorization model, patience increases monotonically  from 2.13 times to 6.23 times as customers type more words. These findings suggest that encouraging or even incentivizing customers to write their inquiries while waiting could significantly increase their patience.

Using an Erlang-A steady-state queueing model \citep{Garnett2002}, we estimate the system’s performance with and without writing in queue, while taking into account their impact on service time and patience. Without allowing writing in queue and keeping the current number of servers,\footnote{We assume that the arrival rate ($\lambda$) is 594.79 customers per hour. The number of ``servers" ($n$) is calculated by the number of concurrency slots: 134.69 servers $\times$ avg.\ concurrency of 3.98, resulting in 536.07 effective servers.} the abandonment rate would increase significantly from 17.2\% to 46.7\% and the probability of waiting will remain as is (close to 1). To maintain current service levels of 17.2\% abandonment without allowing  writing in queue, 22 additional servers would be required. By contrast, allowing queue writing could achieve better performance with fewer resources; for example, adding 14 servers reduces the probability of waiting to 55\% and the probability of abandonment to 3\%, while adding 16 servers reduces the probability of waiting to 40\% and the abandonment to 1.8\%. 

\begin{figure}[!htb]
\centering
\includegraphics[width=0.5\textwidth]{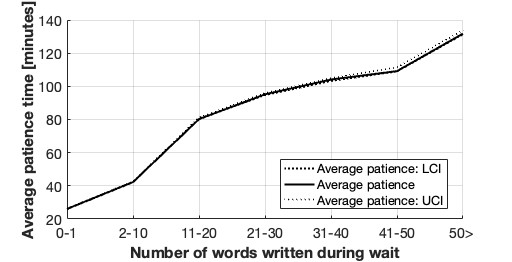}
\caption{Patience as a Function of the Number of Words Written during Waiting in Queue. XYZ Dataset. (Confidence Intervals are Computed Using Bootstrapping.)}
\label{fig:Patience&Words}
\end{figure}

As a robustness check, we repeat the basic EM model (from Section \ref{sec:EM_model}), splitting conversations into groups based on the number of words the customer wrote during their waiting. We then estimated the average patience in each group using our original EM algorithm. The results, presented in Figure \ref{fig:Patience&Words}, align closely with the patterns observed in Table \ref{tbl:coef}. Specifically, as customers write more words while waiting, thereby investing greater effort, their patience substantially increases. For instance, customers who write 18.2 words---the average in the database---exhibit patience that is over three times larger than that of customers who write one word or fewer. 

Additionally, we identify several other factors associated with patience across both models. Customers who express negative sentiment in their queue messages are less patient compared to those with neutral sentiment. Customers accessing the service via the web also exhibit lower patience levels than do those using a mobile app. Returning customers, on the other hand, tend to be more patient than new customers are. 

\subsection{Discussion: Enhancing Customer Satisfaction and Reducing Abandonment}
\label{sec:BoostingExpiriance}




Enhancing customer satisfaction begins with ensuring service is initiated before customers reach the limit of their patience. To achieve this, we propose two complementary strategies: (1) increasing the time customers are willing to wait in the queue (i.e., increasing their patience) and (2) optimizing queue management based on predicted customer patience levels. 

\subsection*{Increasing Customer Patience}

A company can encourage customers to remain in the queue longer by adopting two approaches: (1) fostering deeper engagement with the service system during the wait and (2) providing accurate wait time estimates. The analysis in Section \ref{sec:EMextension-Model}, supports the first approach, while prior research on delay announcements supports the second (see review by \citealp{Ibrahim2018DelayAnnReview} and references therein). 

Encouraging customers to write messages while waiting is a simple yet effective method to enhance engagement and patience. For example, prompting customers with questions like ``While waiting for a live agent, please describe how we can assist you today?" can lead to increased patience, as suggested by the results in Table \ref{tbl:coef}. Implementing AI-powered chatbots to encourage detailed descriptions or handle simple inquiries presents another opportunity to boost engagement. 
However, given the mixed opinions on AI chatbots' impact on customer satisfaction \citep{ruan_when_2022, ashfaq_i_2020}, careful design and deployment are essential, enhance engagement and improve overall customer satisfaction.

Providing wait-time announcements is another effective strategy to increase patience. Uncertainty about wait times often heightens anxiety, whereas predictable waits are generally more tolerable \citep{Maister2005TheLines}. By offering estimated wait times and timely updates, companies can encourage customers to remain in the queue  \citep{huang_refined_2017,Qiuping2017DelayAnnouncements,Qiuping2022RideSharing}. However,  underestimating wait times can have the opposite effect, leading to frustration and reduced satisfaction \citep{whiting_closing_2009}. 
Silent abandonment presents an additional challenge for delay announcement accuracy, as systems may mistakenly assume customers are still waiting when they have already left. 
The extended EM patience estimation model developed in this study highlights the value of incorporating customer features and in-queue message content to better predict abandonment tendencies. Future research can build on these insights by leveraging the EM model to improve delay estimators that account for silent abandonment, ultimately enhancing the customer experience.

\subsection*{Optimizing Queue Management}
The second strategy involves utilizing patience predictions to optimize queue prioritization. Previous research  has demonstrated that customer patience is an important factor in determining customer priority in multiclass queueing systems \citep{Atar2010cMuTheta,wu_service_2019}. By analyzing customer characteristics upon queue entry and updating patience estimates dynamically based on new messages, our model facilitates the development of \emph{personalized patience estimation} for each customer. This approach enables companies to tailor their prioritization policies to the dynamic conditions of customer patience.

For instance, a company can leverage our model and prioritize customers with lower predicted patience---a policy known as a the \emph{least-patient-first} policy. This policy refines the traditional first-come-first-served method  by identifying and serving more impatient customers earlier, thereby reducing abandonment rates. Originally proposed by \citet{Mandelbaum2017}, this policy has been successfully implemented in call centers \citep{adan_first-come_2019, wu_service_2019}.

Our model enriches previous research by categorizing customer patience dynamically using diverse attributes, such as platform, return status, sentiment, and dynamic queue engagement.
This multidimensional characterization of patience enables more comprehensive queue optimization strategies, such as those suggested by \citet{Bassamboo2016Scheduling}, which account for general patience distributions. By tailoring queue management to individual patience profiles, businesses can enhance service efficiency, improve customer satisfaction, and minimize frustration.

\section{Conclusion}
\label{sec:Conclusion}

In this article, we identified and defined the phenomenon of silent abandonment as an important source of uncertainty in contact centers. We laid the groundwork for identifying silent abandonment in contact centers by training classification models  incorporating text analysis of customer and agent messages, utilizing features derived from NLP techniques. Silent abandonment appears to be widespread in the industry, with 22\% of an average company conversations classified as uncertain silent abandonment based on their metadata. Detailed analysis of the XYZ dataset revealed that 12.6\% of all conversations (or 51.8\% of uSab cases) are indeed instances of silent abandonment, effectively tripling the rate of abandonment previously considered in the company’s performance metrics. Additionally, we demonstrated that silent abandonment increases agent workload and customer delays. 
These findings underscore the necessity of revising performance measurement calculations and operational methods to account for silent abandonment.
Future improvements to these classification models could involve fine-tuning a large language model (LLM), such as BERT \citep{devlin2019bert}. 

We analyzed the implications of silent abandonment on customer-patience estimation and demonstrated the necessity of treating silent abandonment as uncertain left-censored observations to achieve accurate estimates. Using an EM algorithm, we estimated customer patience in the XYZ contact center to be 81.1 minutes. This extended patience likely stems from allowing customers to engage with the service system by writing inquiries while waiting, a design choice shown in Section \ref{sec:patience_coef} to triple customer patience.

We introduced an extended EM algorithm designed to identify such associations between various factors and customer patience. This EM framework could be applied to other environments where silent abandonment is prevalent and resources are scarce, such as interactive AI systems with a human in the loop and healthcare settings like emergency rooms. However, the EM algorithm has certain limitations, including that it assumes that patience is distributed exponentially and that offered wait time and patience are independent. Furthermore, the extended EM  algorithm identifies correlations rather than causal relationships. To validate the causal impact of allowing customers to write while waiting in the queue, we recommend conducting online A/B tests, comparing outcomes for customers who are permitted to write during their wait to those who are not. We leave this experiment for future research.

From a behavioral perspective, we want to highlight the potential for agents to exploit silent abandonment to appear busy (by not closing these conversations) while actually using the time for rest. 
Companies want to prevent such strategic behavior but should proceed carefully to avoid situations where a long-waiting customer conversation is prematurely terminated. For example, it is possible that the customer simply did not notice that the agent had finally answered. Hence, finding technological solutions to mitigate capacity loss, like the ones we suggested in Section \ref{sec:avoid_Sab}, 
is important. Companies should also revisit policies for automatically closing conversations, as discussed in \cite{castellanos2024closing}, to better account for silent abandonment without prematurely terminating conversations that may still require attention.

To conclude, we believe that the phenomenon of silent abandonment extends beyond the scope discussed in this paper. It warrants further exploration, both mathematically and behaviorally, in the context of text-based services and beyond.

%
%
%



\bibliographystyle{informs2014} 
\bibliography{Mendeley1.bib, Abandonment_ref.bib} 

\ECSwitch

\ECDisclaimer


\section{Summary Statistics} \label{app:summary_stats}

The XYZ dataset originally included 332,978 conversations. After excluding 1,391 observations (0.4\% of the conversations) due to unrealistic or incomplete information---for example, instances where the closure time was recorded before the last message was sent or where no events were registered in the conversation---we analyzed a final dataset of 331,587 conversations.

Table \ref{tbl:general_stat} provides summary statistics for the dataset as a whole as well as broken down by conversation type (Sr, Kab, and uSab, with the latter further divided into Sr1 and Sab). The table includes metrics such as customer wait time for agent assignment; customer response time (RT), defined as the duration between an agent message and the customer's response; agent response time, defined as the duration between a customer message and the agent's response; the time elapsed between the last message sent and conversation closure; the number of messages and words written by each party; and other relevant features.

\begin{figure}[htb]
\centering
\subfigure[Conversation Start Day by Customer Type]{
\includegraphics[width=0.45\textwidth]{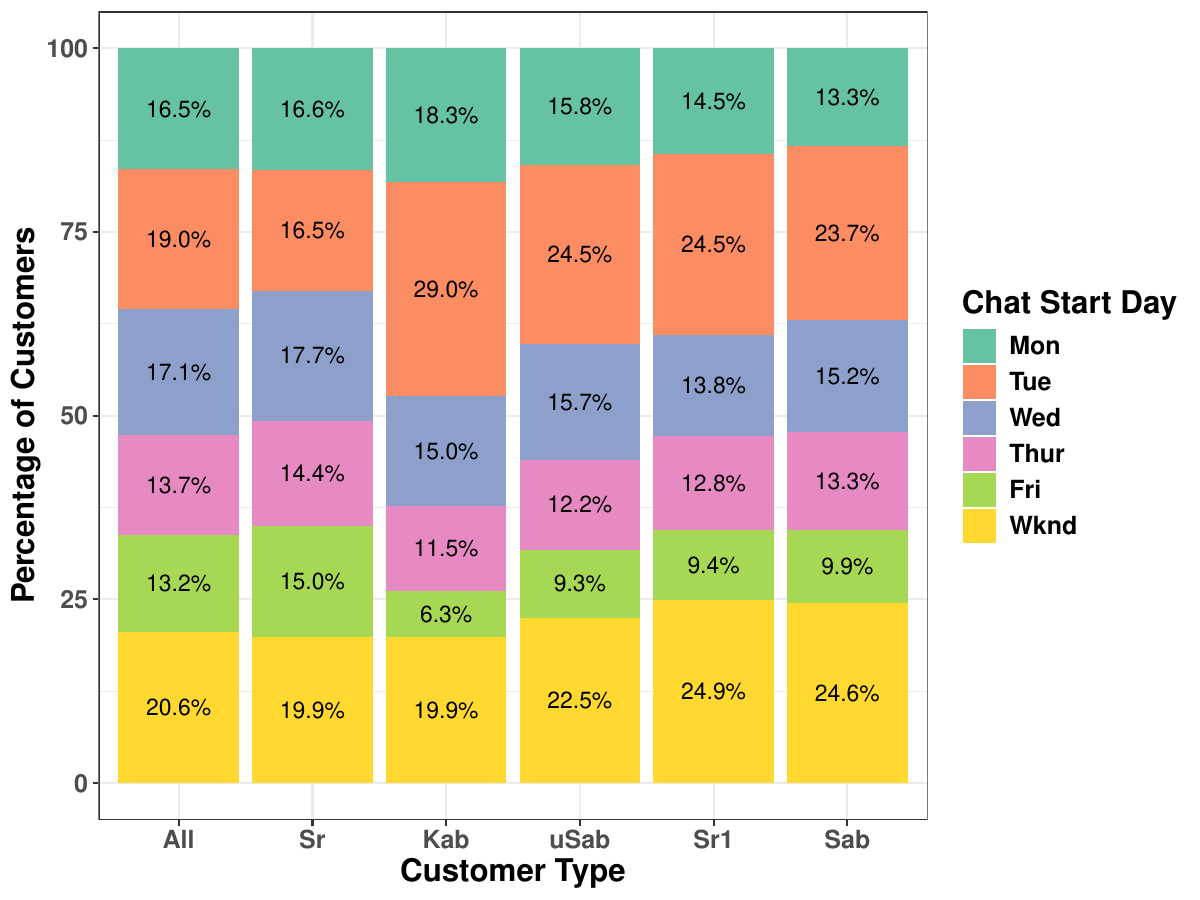} \label{fig:ChatStart_a}
} 
\subfigure[Conversation Start Hour by Customer Type]{
\includegraphics[width=0.45\textwidth]{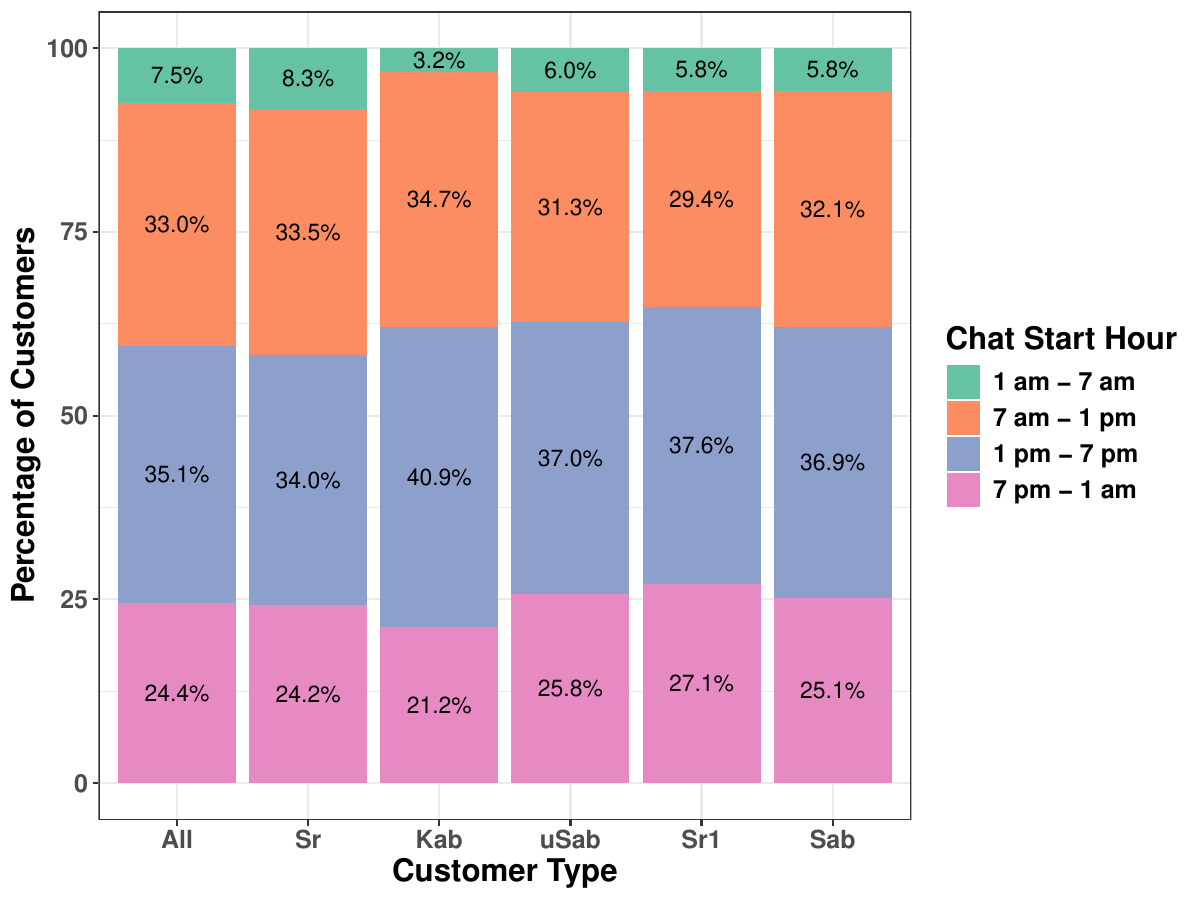}
\label{fig:ChatStart_b}
}
\caption{Conversation Start Day and Hour by Customer Type (Percentages)}
\label{fig:ChatStart}
\end{figure}

Figure \ref{fig:ChatStart} illustrates the distribution of conversations by day of the week and time of day for each customer type. Notably, Sab conversations occur across all days of the week and at all times of day.
 

\begin{table}[!htb]
 \centering
 \caption{General Statistics}
  \label{tbl:general_stat}

  \begin{scriptsize}
  \begin{tabular}{lcccccccc} 
  \toprule
Cust.\ &\multicolumn{1}{c}{\% of}&\# agent msg.\ &\# agent words&\multicolumn{5}{c}{Agent RT in minutes [Mean (SD)]}\\  \cmidrule{5-9}
type&\multicolumn{1}{c}{conv.\ }&\multicolumn{1}{c}{Mean (SD)}&\multicolumn{1}{c}{ per msg. }&\multicolumn{1}{c}{All}&\multicolumn{1}{c}{Turn 1}&\multicolumn{1}{c}{Turn 2}&\multicolumn{1}{c}{Turn 3}&\multicolumn{1}{c}{Turn 4}\\ \midrule
All&&~9.7 (8.98)&23.0 (22.74)&~~3.6 (15.16)&2.2 (5.3)&~~3.8 (15.1)&~~5.2 (18.4)&~~5.2 (18.9)\\
Sr&70.6\%~~&11.8 (9.36)&22.2 (22.16)&~~3.0 (12.74)&1.9 (3.8)&~~2.1 (~8.2)&~~3.0 (10.4)&~~3.5 (13.3)\\
Kab&~~5.1\%~~&\multicolumn{1}{c}{--}&\multicolumn{1}{c}{--}&\multicolumn{1}{c}{--}&\multicolumn{1}{c}{--}&\multicolumn{1}{c}{--}&\multicolumn{1}{c}{--}&\multicolumn{1}{c}{--}\\
uSab&24.4\%~~&~3.3 (2.18)&31.4 (27.02)&10.8 (30.36)&3.1 (8.3)&10.2 (28.0)&16.5 (37.5)&18.2 (40.4)\\
~-Sr1&48.2\%$^*$&~4.3 (2.45)&34.5 (29.81)&12.3 (33.30)&3.4 (8.7)&10.6 (28.5)&17.3 (39.2)&18.4 (41.1)\\
~-Sab&51.8\%$^*$&~2.3 (1.29)&26.2 (20.29)&~~8.4 (24.68)&2.8 (5.3)&~~9.8 (27.8)&15.5 (35.4)&17.7 (37.7)\\ 
\bottomrule
\multicolumn{9}{l}{$^*$Percentage of uSab. Estimated using SVM with threshold of 0.47 (\S\ref{sec:prob_abnd_WQ}).}\\
\multicolumn{9}{l}{Note: Standard deviations are provided in parentheses.}\\
 \end{tabular}

\vspace*{3mm}

\begin{tabular}{lcccccccccccc} \toprule
Cust.\ &\multicolumn{2}{c}{\# of cust.\ messages}&&\multicolumn{2}{c}{Cust.\ words per msg. }&&\multicolumn{2}{c}{Cust.\ RT per msg.\ [min]} && \multicolumn{2}{c}{Cust.\ chars.\ per msg.\ }  \\ \cmidrule{2-3} \cmidrule{5-6} \cmidrule{8-9} \cmidrule{11-12}
type&All&In-queue&&All&In-queue&&All&First turn$^1$ &&All&In-queue  \\ \midrule
All&6.0 (6.5)&1.2 (0.5)&&13.1 (16.0)&18.2 (24.6)&&1.0 (3.6)&1.4 (5.2) &&  377.9 (513.6) &105.6 (125.9)\\
Sr&7.9 (6.7)&1.2 (0.5)&&12.9 (15.7)&18.2 (24.9)&&1.0 (3.5)&1.4 (5.2) &&  495.7 (565.4) &110.4 (127.1)\\
Kab&1.2 (0.5)&1.2 (0.5)&&16.4 (25.1)&16.4 (25.1)&&2.1 (5.3)&\multicolumn{1}{c}{--} &&  ~93.0 (128.5) & ~93.0 (128.5)\\
uSab&1.1 (0.4)&1.1 (0.4)&&18.2 (20.5)&18.5 (23.6)&&1.1 (4.3)&\multicolumn{1}{c}{--} &&  ~94.3 (120.7) & ~94.3 (120.7)\\
~-Sr1&1.1 (0.4)&1.1 (0.4)&&22.0 (22.6)&22.0 (26.5)&&1.0 (4.1)&\multicolumn{1}{c}{--}&&  115.3 (136.2) &115.3 (136.1)\\
~-Sab&1.1 (0.4)&1.1 (0.4)&&15.3 (18.2)&15.3 (20.7)&&1.2 (4.7)&\multicolumn{1}{c}{--}&&  ~79.0 (104.9) & ~79.0 (105.1)\\ \bottomrule
\multicolumn{12}{l}{Note: Standard deviations are provided in parentheses.}\\
\multicolumn{12}{l}{$^1$First turn after agent assignment.}

\end{tabular}
\vspace*{3mm}

\begin{tabular}{lccccc} \toprule
Customer &Wait time&Service time&Closure time&Concurrency time&Total time\\ 
type&\multicolumn{1}{c}{Mean (SD)}&\multicolumn{1}{c}{Mean (SD)}&\multicolumn{1}{c}{Mean (SD)}&\multicolumn{1}{c}{Mean (SD)}&\multicolumn{1}{c}{Mean (SD)}\\ \midrule
All&~8.3 (18.28)&46.3 (63.56)&65.9 (75.08)&112.3 (92.73)&120.6 (95.80)\\
Sr&~5.3 (11.59)&53.8 (65.36)&59.2 (73.51)&113.0 (95.61)&118.3 (97.04)\\
Kab&~7.3 (~~8.64)&\multicolumn{1}{c}{--}&\multicolumn{1}{c}{--}&\multicolumn{1}{c}{--}&~~~7.3 (~~8.64)\\
uSab&17.3 (29.29)&34.3 (58.76)&99.1 (73.11)&133.4 (75.85)&150.7 (82.36)\\
~-Sr1&18.9 (30.26)&53.7 (71.36)&95.0 (74.99)&148.7 (77.00)&167.5 (81.86)\\
~-Sab&19.4 (31.12)&20.1 (40.65)&113.6 (65.88)&133.7 (63.62)&153.1 (71.29)\\ \bottomrule
\multicolumn{6}{l}{Note: Times are measured in minutes.}
 \end{tabular}

\vspace*{3mm}

\begin{tabular}{lccccccccccccccc} \toprule
Cust.\ &\multicolumn{4}{c}{\% terminated by}&&\multicolumn{3}{c}{Platform (\%)}&& \multicolumn{2}{c}{ Returning cust.\ (\%) } && \multicolumn{3}{c}{Cust.\ sent.\ in queue (\%)} \\ \cmidrule{2-5} \cmidrule{7-9} \cmidrule{11-12} \cmidrule{14-16}
type&System&Agent&Customer&Manager&&App&Web&Other && Yes & No && Pos.\ & Neg.\ & Neut.\  \\ \midrule
All&32.0\%&47.8\%&~19.9\%&0.3\%&&56.5\%&42.7\%&0.9\% && 41.7\% & 58.3\% && 4.5\% & 14.7\% & 80.8\%\\
Sr&28.0\%&53.1\%&~18.5\%&0.4\%&&57.8\%&42.0\%&0.2\% && 43.2\% & 56.8\% && 3.9\% & 15.4\% & 80.7\% \\
Kab&~~~0.0\%&~~0.0\%&100.0\%&0.0\%&&28.7\%&71.3\%&0.0\% && 27.2\% & 72.8\% && 4.8\% & 13.6\% & 81.7\% \\
uSab&50.3\%&42.2\%&~~7.2\%&0.3\%&&58.4\%&38.8\%&2.8\% && 40.4\% & 59.6\% && 6.0\% & 13.1\% & 81.5\% \\
~-Sr1&47.4\%&50.6\%&~~~~1.8\%&0.3\%&&59.9\%&37.1\%&3.0\% && 46.9\% & 53.1\% && 4.4\% & 14.4\% & 81.2\% \\
~-Sab&58.2\%&37.7\%&~~~~4.0\%&0.2\%&&58.4\%&38.8\%&2.9\% && 35.0\% & 65.0\% && 6.3\% & 12.0\% & 81.7\%\\ \bottomrule
 \end{tabular}

  \end{scriptsize}


\end{table}

\section{Classification Models} \label{app:ClassMod}
We provide here more details on the two best classification models described in Section \ref{sec:prob_abnd_WQ}: the SVM model and the classification tree. For both models, we selected the features for the final models using a wrapper method with recursive feature elimination \citep{guyon2002gene}.
\subsection{Support Vector Machines} \label{app:SVM}
The fitted model has 259 support vectors. The variables included in this model are:
\begin{enumerate}
    \item AgentChars: the number of characters written by the agent in the conversation.
    \item AgentDuration: the time it takes the agent to write their messages.
    \item QueueTime: the time the customer waits for agent assignment in the queue.
    \item TotalDuration: the time from agent assignment until manual closure of the conversation by the agent or automatic closure by the system.
    \item wordag46, wordag1, wordag20, wordag31: specific words written by the agent during the conversation. 
    \item wordcust1: a specific word written by the customer in the initial inquiry.
\end{enumerate}   
We provide only coded words due to  privacy concerns of the company providing the dataset.

\subsection{Classification Tree}

\begin{figure}[!htb] 
\centering

\includegraphics[width=.7\textwidth]{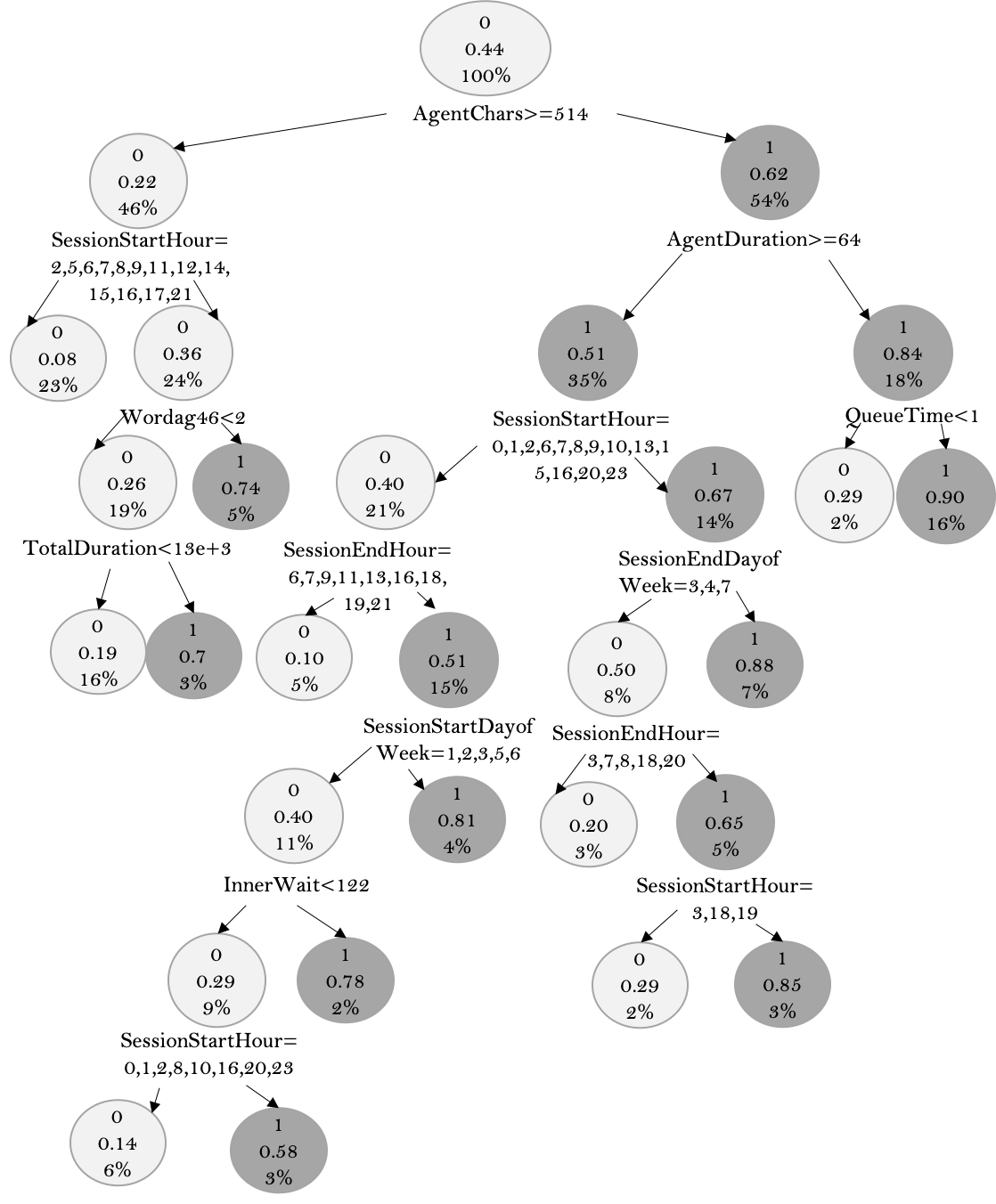}
\caption{Classification Tree for the Probability of a Conversation Being a Silent Abandonment. Splitting Variable on the Bottom of the Nodes. Nodes Show 1 for Silent Abandonment (Grey) and 0 for Served-in-one-exchange (Light Grey); Probability of Obtaining that Classification; Percentage of the Data that Falls into That Node. }
\label{fig:ClasfTree}
\end{figure}

The fitted classification tree is presented in Figure \ref{fig:ClasfTree}. The variables included in this model are:
\begin{enumerate}
\item AgentChars, AgentDuration, QueueTime, TotalDuration, wordag46: see \ref{app:SVM}.
\item SessionStartHour: the hour that the customer arrived to the system.
\item SessionStartDayofWeek: the day of the week on which the customer arrived to the system.
\item InnerWait: the time the customer waits for the agent's reply during service (i.e., after assignment to the agent).
\item SessionEndHour: the hour that the conversation was closed; the conversation may be closed manually by the agent (usually within a few hours of no customer reply) or automatically by the system (after a threshold time has passed).
\item SessionEndDayofWeek: the day of the week on which the conversation is closed.
\end{enumerate}   


\section{Classification Models That Use Only Information in Queue} \label{app:ClassModQueue}
In the discussion of Section \ref{sec:avoid_Sab}, we suggested updating the company operation based on suspected silent abandonment. We present in Figure \ref{fig:ROCQueue} and Table \ref{tbl:AUCQueue} ROC curves of our initial attempts to fit a classification model that can predict silent abandonment based on data available while waiting.  The variables selected using recursive feature elimination are Queue time, Number of words written in messages while waiting in queue, Returning customer indicator, Platform indicator, Chat starting hour and day, 
Sentiment, and five specific words used by customers.
The SVM classifier has the best results and includes 336 support vectors. The areas under the ROC curves for the models (Table \ref{tbl:AUCQueue}) are not high, which suggests that further research to improve is needed.
\begin{table}[!htb]
   \begin{minipage}{.5\textwidth}
    \centering
    \includegraphics[width=0.75\textwidth]{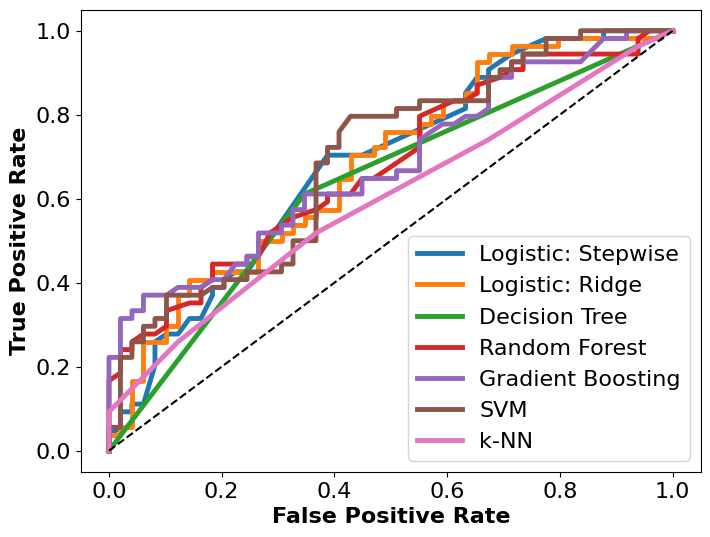}%
    \captionof{figure}{ROC Curve on Test Dataset}
    \label{fig:ROCQueue}
  \end{minipage}
  \begin{minipage}{0.3\textwidth}
    \caption{Area under the ROC}  
    \label{tbl:AUCQueue}
    \centering
    \begin{small}
    \centering
    \begin{tabular}{lc} 
    \toprule
    Model & AUC\\ \midrule 
    SVM & 0.69\\
    Logistic Regression: Stepwise & 0.68\\
    Logistic Regression: Ridge & 0.68\\
    Gradient Boosting & 0.68\\
    Random Forest  & 0.67\\
    Decision Tree & 0.63\\     
    k-NN & 0.60\\     
    \bottomrule 
    \end{tabular}
    \end{small}
  \end{minipage}\hfill
\end{table}


\section{EM Algorithm: Proof and Explanation} \label{app:EMproofs}
The log of the likelihood in Eq.\ \eqref{eq:Likelihood} is
\begin{equation} \label{eq:loglikelihood}
\begin{split}l(D,\theta,q,\gamma)=\sum_{i=1}^{n}\left\{ \left(1-\Delta_{i}\right)\left(\log\gamma-\gamma U_{i}-\theta U_{i}\right)\right\} +\sum_{i=1}^{n}\left\{ \left(\Delta_{i}Y_{i}\right)\left[\log\theta-\theta U_{i}-\gamma U_{i}+\log(q)\right]\right\} \\
+\sum_{i=1}^{n}\left\{ \left(\Delta_{i}(1-Y_{i})\right)\left[\log(1-q)+\log(1-e^{-\theta U_{i}})+\log\gamma-\gamma U_{i}\right]\right\},
\end{split}
\end{equation}
wherein, if the data is complete, the possible classes for  conversation $i$ are $C_{1}^{i}=1-\Delta_{i}$, $C_{2}^{i}  =Y_{i}\Delta_{i}$, and $C_{3}^{i}=(1-Y_{i})\Delta_{i}$. Therefore, the log-likelihood in Eq.\ \eqref{eq:loglikelihood} when the data is complete can be written as
\begin{equation} \label{eq:loglikelihoodComplete}
\begin{split}l(D,\theta,q,\gamma)=\sum_{i=1}^{n}\left\{ \left(C_{1}^{i}\right)\left(\log\gamma-\gamma U_{i}-\theta U_{i}\right)\right\} +\sum_{i=1}^{n}\left\{ \left(C_{2}^{i}\right)\left[\log\theta-\theta U_{i}-\gamma U_{i}+\log(q)\right]\right\} \\
+\sum_{i=1}^{n}\left\{ \left(C_{3}^{i})\right)\left[\log(1-q)+\log(1-e^{-\theta U_{i}})+\log\gamma-\gamma U_{i}\right]\right\}.
\end{split}
\end{equation}

In the case of missing data in $\Delta$, we cannot maximize the log-likelihood in Equation \eqref{eq:loglikelihoodComplete} because some observations might belong to either class $C_{1}=1$ or $C_{3}=1$. To solve this problem, we use the expectation-maximization (EM) algorithm. It calculates starting parameters with random starting weights for conversation classes (see Algorithm \ref{EM}) and then iterates between two steps---the expectation and the maximization steps---until convergence. 
(The convergence criterion is given in Algorithm \ref{EM}.) 



\subsection{Expectation Step, Proof of Theorem \ref{Theorem1}} 
\label{app:EMproofs-ExpectationStep}
In the $t$th iteration, the expectation step consists of finding a surrogate function that is a lower bound on the log-likelihood in Eq.~\eqref{eq:loglikelihoodComplete} but is tangent to the log-likelihood at $(\widehat{\theta^{(t)}},\widehat{q^{(t)}},\widehat{\gamma^{(t)}})$, the vector of the parameters of the latest iteration, $t-1$. We achieve this by computing the expectation given what we know up to the $t$th iteration, that is, the ($t-1$)th estimations of the parameters $(\widehat{\theta^{(t)}},\widehat{q^{(t)}},\widehat{\gamma^{(t)}})$ and the data that is complete. Formally,
\begin{equation} \label{eq:expectLogLikelihood}
E\left[l(D,\theta,\gamma,q)\mid U_{i},M^{i},\widehat{\theta^{(t)}},\widehat{\gamma^{(t)}},\widehat{q^{(t)}}\right],
\end{equation}
where $M$ is defined as in Section \ref{subsec:ClassesMissingDa}.

When $M^i=2$, the data is complete, implying that for that conversation $C^i_{2}=1$. Therefore, we can compute Eq.\ \eqref{eq:expectLogLikelihood} for such $i$th observations as follows:
\begin{align*}
E\left[C_{2}^{i}=1\mid U_{i},M^{i},\widehat{\theta^{(t)}},\widehat{q^{(t)}},\widehat{\gamma^{(t)}}\right]=E\left[\Delta_{i}Y_{i}\mid U_{i},M^{i},\widehat{\theta^{(t)}},\widehat{q^{(t)}},\widehat{\gamma^{(t)}}\right] & =1_{\{M^{i}=2\}}+\left(0\right)\left(1_{\{M^{i}\neq2\}}\right).
\end{align*}

We define $\widehat{C_{2,t}^{i}}$ as the probability that the $i$th conversation is a known abandonment in the $t$th iteration. This is exactly what the previous function represents. Therefore, 
\[
\widehat{C_{2,t}^{i}}=1_{\{M^{i}=2\}}.
\]

When $M^i=0$, there is missing data. This implies that for conversation $i$ either $C^i_{1}=1$ or $C^i_{3}=1$. We first compute Eq.\ \eqref{eq:expectLogLikelihood} by conditioning on knowing that the $i$th observation is a member of class $C_{3}=1$:
\begin{align*}
E\left[C_{3}^{i}=1\mid U_{i},M^i,\widehat{\theta^{(t)}},\widehat{q^{(t)}},\widehat{\gamma^{(t)}}\right] & =E\left[\Delta_{i}\left(1-Y_{i}\right)\mid U_{i},M^{i},\widehat{\theta^{(t)}},\widehat{q^{(t)}},\widehat{\gamma^{(t)}}\right]\\
 & =\left(1_{\{M_{i}=0\}}\right)\textit{Pr}\left\{ \Delta_{i}=1\mid U_{i},M^{i}=0,\widehat{\theta^{(t)}},\widehat{q^{(t)}},\widehat{\gamma^{(t)}}\right\} +\left(0\right)\left(1_{\{M_{i}\neq0\}}\right)\\
 & =\left(1_{\{M_{i}=0\}}\right)\textit{Pr}\left\{ \tau_{i}\leq W_{i}\mid U_i = W_i,\widehat{\theta^{(t)}},\widehat{\gamma^{(t)}}\right\} \\
 & =\left(1_{\{M_{i}=0\}}\right)\textit{Pr}\left\{ \tau_{i}\leq U_{i}\mid U_{i},\widehat{\theta^{(t)}}\right\} \\
 & =1_{\{M_{i}=0\}}\left(1-e^{-\widehat{\theta^{(t)}}U_{i}}\right).
\end{align*}

The first equality follows since a customer $i$ for which
$M^i=0$ does not give an indicator when abandoning the queue, that is, $Y_i=0$, and by the independence of $\Delta_i$ and $Y_i$. The second equality follows since a customer $i$ that belongs to class $C_{3}=1$  must have $M^i=0$, and clearly that customer abandoned. Hence, $\Delta_i=1$ by definition, which formally means that $\tau_i\leq W_i$. Additionally, when $ M^{i}=0$, the observed time $U_i$ is their wait time $W_i$; hence, the third equality follows.
The fourth equality is implied from the third.
The fifth equality follows since the fourth equality represents the CDF $F$
of patience time $\tau$ that has an exponential distribution by Assumption \ref{assumption:1}.

We define $\widehat{C_{3,t}^{i}}$ as the estimated probability that a customer $i$ is a silent abandonment in the $t$th iteration, which by the above calculation is
\[
\widehat{C_{3,t}^{i}}=1_{\{M^{i}=0\}}\left(1-e^{-\widehat{\theta^{(t)}}U_{i}}\right).
\]

Note that in the case where $M^{i}=0$ for conversation $i$, we need to consider also that the customer might belong to $C_{1}^{i}=1$. In addition, we have another group that belongs to class $C_{1}=1$, namely the conversations where $M^{i}=1$. For them, the data is complete. So, the computation of Eq.\ \eqref{eq:expectLogLikelihood} for a customer $i$ that belongs to class $C_{1}^{i}=1$ is as follows:
\begin{align*}
E\left[C_{1}^{i}=1\mid U_{i},M^{i},\widehat{\theta^{(t)}},\widehat{q^{(t)}},\widehat{\gamma^{(t)}}\right] & =E\left[1-\Delta_{i}\mid U_{i},M^{i},\widehat{\theta^{(t)}},\widehat{q^{(t)}},\widehat{\gamma^{(t)}}\right]\\
 & =\left(1_{\{M^{i}=0\}}\right)\textit{Pr}\left\{ \Delta_{i}=0\mid U_{i},M^{i}=0,\widehat{\theta^{(t)}},\widehat{q^{(t)}},\widehat{\gamma^{(t)}}\right\} +1_{\{M^{i}=1\}}\\
 & =\left(1-\widehat{C_{3,t}^{i}}\right)1_{\{M^{i}=0\}}+1_{\{M^{i}=1\}}.
\end{align*}
The second term on the right-hand side of the second equality ($1_{\{M^{i}=1\}}$) follows since for the customers that are classified as $M=1$, the data is complete. The first term follows since some of the customers in $M=0$ belong as well to $C_1=1$, and these customers are the ones who do not abandon, that is, $\Delta=0$.
The third equality follows since the probability of a  customer $i$ in $M^{i}=0$ being served (not abandoned) is exactly the complement of $\widehat{C_{3,t}^{i}}$.

We define $\widehat{C_{1,t}^{i}}$ as the probability that a customer $i$ is a served customer, which by the above computation is
\begin{align*}
\widehat{C_{1,t}^{i}} & =\left(1-\widehat{C_{3,t}^{i}}\right)1_{\{M^{i}=0\}}+1_{\{M^{i}=1\}}.
\end{align*}


Finally, we can rewrite our log-likelihood, Eq.~\eqref{eq:loglikelihood}, in the $t$th iteration with missing data as Eq.~\eqref{eq:loglikelihoodMisg}. This is exactly the surrogate function that is a lower bound on the log-likelihood: the E-step in Algorithm \ref{EM}. 

\begin{flushright}
$\blacksquare$
\par\end{flushright}

\subsection{Maximization Step, Proof of Theorem \ref{Theorem2}} \label{app:EMproofs-MaximizationStep}
In the $t$th iteration of the maximization step, the parameters $(\widehat{\theta^{(t+1)}},\widehat{q^{(t+1)}},\widehat{\gamma^{(t+1)}})$ are found to be the maximizers of the surrogate function defined in Eq.~\eqref{eq:loglikelihood}. We obtain the parameters $(\widehat{\theta^{(t+1)}},\widehat{q^{(t+1)}},\widehat{\gamma^{(t+1)}})$, where the partial derivatives of the surrogate function \eqref{eq:loglikelihoodMisg} are equal to zero. 

The partial derivative with respect to $q$ is
\[
\begin{split}\frac{\partial\ell}{\partial q} & =\end{split}
\left(\frac{1}{q}\right)\sum_{i=1}^{n}\widehat{C_{2,t}^{i}}-\left(\frac{1}{1-q}\right)\sum_{i=1}^{n}\widehat{C_{3,t}^{i}},
\]
which yields
\[
\widehat{q^{(t+1)}}=\left\{ \sum_{i=1}^{n}\widehat{C_{2,t}^{i}}\right\} \left\{ \sum_{i=1}^{n}\left(1-\widehat{C_{1,t}^{i}}\right)\right\} ^{-1}.
\]


The partial derivative with respect to $\gamma$ is
\[
\frac{\partial\ell}{\partial\gamma}=\frac{1}{\gamma}\sum_{i=1}^{n}\left(1-\widehat{C_{2,t}^{i}}\right)-\sum_{i=1}^{n}U_{i},
\]
which yields
\[
\widehat{\gamma^{(t+1)}}=\left\{ \sum_{i=1}^{n}\left(1-\widehat{C_{2,t}^{i}}\right)\right\} \left\{ \sum_{i=1}^{n}U_{i}\right\} ^{-1}.
\]


The partial derivative with respect to $\theta$ is

\begin{align*}
\frac{\partial\ell}{\partial\theta} & =\sum_{i=1}^{n}\left(\widehat{C_{1,t}^{i}}\right)\left(-U_{i}\right)+\sum_{i=1}^{n}\left(\widehat{C_{2,t}^{i}}\right)\left(-U_{i}+\frac{1}{\theta}\right)+\sum_{i=1}^{n}\widehat{C_{3,t}^{i}}\frac{U_{i}e^{-\theta U_{i}}}{1-e^{-\theta U_{i}}}\\
 & =\sum_{i=1}^{n}\left(U_{i}\right)\left(-\widehat{C_{1,t}^{i}}-\widehat{C_{2,t}^{i}}\right)+\sum_{i=1}^{n}\left(\frac{\widehat{C_{2,t}^{i}}}{\theta}\right)+\sum_{i=1}^{n}\widehat{C_{3,t}^{i}}\frac{U_{i}e^{-\theta U_{i}}}{1-e^{-\theta U_{i}}}\\
 & =\sum_{i=1}^{n}\left(U_{i}\right)\left(\widehat{C_{3,t}^{i}}-1\right)+\sum_{i=1}^{n}\left(\frac{\widehat{C_{2,t}^{i}}}{\theta}\right)+\sum_{i=1}^{n}\widehat{C_{3,t}^{i}}\frac{U_{i}e^{-\theta U_{i}}}{1-e^{-\theta U_{i}}}.
\end{align*}
The second equality follows from simplifying the terms. The third equality follows from the relation $-\widehat{C_{1,t}^{i}}-\widehat{C_{2,t}^{i}}-\widehat{C_{3,t}^{i}}=-1$ for the $i$th customer in the $t$th iteration. 


Finally, we set the derivative to zero and multiply by $\widehat{\theta^{(t+1)}}$: 
\[
\widehat{\theta^{(t+1)}}\left\{ \sum_{i=1}^{n}\left(\widehat{C_{3,t}^{i}}-1\right)U_{i}\right\} +\sum_{i=1}^{n}\widehat{C_{2,t}^{i}}+\widehat{\theta^{(t+1)}}\left\{ \sum_{i=1}^{n}\widehat{C_{3,t}^{i}}\frac{U_{i}e^{-\widehat{\theta^{(t+1)}}U_{i}}}{1-e^{-\widehat{\theta^{(t+1)}}U_{i}}}\right\} =0.
\]

To show that the solutions are indeed maximizers, we calculate the Hessian of Eq.\  \eqref{eq:loglikelihoodComplete} and show that it is a negative-definite matrix, as follows:
\[
\begin{aligned}\frac{\partial^{2}\ell}{\partial q^{2}}= & -\frac{1}{q}\sum_{i=1}^{n}\widehat{C_{1,t}^{i}}-\frac{1}{\left(1-q\right)^{2}}\sum_{i=1}^{n}\widehat{C_{3,t}^{i}},\\
\frac{\partial^{2}\ell}{\partial\gamma^{2}}= & -\frac{1}{\gamma^{2}}\sum_{i=1}^{n}\left(1-\widehat{C_{2,t}^{i}}\right),\\
\frac{\partial^{2}\ell}{\partial\theta^{2}}= & -\frac{1}{\theta^{2}}\sum_{i=1}^{n}\left(\widehat{C_{2,t}^{i}}\right)-\sum_{i=1}^{n}\left(\widehat{C_{3,t}^{i}}\right)\frac{U_{i}^{2}e^{-\theta U_{i}}}{\left(1-e^{-\theta U_{i}}\right)^{2}},
\end{aligned}
\]
where $q\in(0,1)$ and $\widehat{C_{j,t}^{i}}\in[0,1]$,
$\forall i,j,t$. We assume that $\gamma>0$, since average virtual
wait time cannot be negative or infinite, and that $\theta>0$, since average customer patience cannot be negative or infinite. Additionally, $\sum_{i=1}^{n}\widehat{C_{1,t}^{j}}>0$ for
$j=1,2,3$ and any $t$, since we assume the system has at least one customer from each of the classes defined in Table \ref{tbl:MissingDtaNotation}. Finally, $\left(1-e^{-\theta U_{i}}\right)\in(0,1]$
since $U_{i}>0$ $\forall i$, that is, because we expect customers to have some patience and for the system to offer at least some virtual wait (at least a small $\epsilon$). Therefore, all the second-order partial derivatives are negative. All the second-order mixed derivatives of Eq.\ \eqref{eq:loglikelihoodComplete} are equal to zero; therefore, its Hessian is a $3\mathrm{x3}$ diagonal matrix where the entries are negative, which means that the Hessian is a negative-definite matrix.
\begin{flushright}
$\blacksquare$
\par\end{flushright}

\subsection{Extension: The EM Algorithm with Patience Covariates.}
\label{app:EMextantion-proofs}
Assume that the patience parameter, $\theta$, is influenced by $k$ variables  $\textbf{X}=[X_{1},...,X_{k}]$ such that  $\theta | \textbf{X}\triangleq e^{-(\beta_{0}+\beta_{1}X_{1}+...+\beta_{k}X_{k})}=e^{-(\beta_{0}+\beta^{T} \textbf{X})}$. Using similar notation to Section \ref{subsec:EM_algorithm}, the observed data now consists of $D \triangleq \{(U_{i},Y_{i},\triangle_{i},\textbf{X}_{i})$, $i=1,...,n\}$. 
We redefine the log-likelihood function with missing data as follows:
%

\begin{equation} \label{eq:log_lik_cov_raw}
\begin{aligned}l(D,\theta,q,\gamma) & =\sum_{i=1}^{n}\left\{ \left(\widehat{C_{1,t}^{i}}\right)\left[-U_{i}e^{-(\beta_{0}+\beta^T\textbf{X}_{i})} + \log\gamma-\gamma U_{i}\right]\right\} \\
 & +\sum_{i=1}^{n}\left\{ \left(\widehat{C_{2,t}^{i}}\right)\left[-\beta_{0}-\beta^T\textbf{X}_{i}- U_{i}e^{-(\beta_{0}+\beta^T\textbf{X}_{i})}-\gamma U_{i}+\log q\right]\right\} \\
 & +\sum_{i=1}^{n}\left\{ \left(\widehat{C_{3,t}^{i}}\right)\left[\log(1-q)+\log\left(1-e^{-U_{i}e^{-(\beta_{0}+\beta^T\textbf{X}_{i})}}\right) +\log\gamma -\gamma U_i\right]\right\},
\end{aligned}
\end{equation}
where
 $\beta\triangleq\begin{bmatrix}\beta_{1} & \cdots & \beta_{k}\end{bmatrix}$ and 
$\textbf{X}_{i}\triangleq\begin{bmatrix}X_{i,1} & \cdots & X_{i,k}\end{bmatrix}$. This likelihood after simplification is Equation \eqref{eq:log_lik_cov}.

The E-Step of the algorithm does not change. In obtaining $\widehat{C_{1,t}^{i}}$, $\widehat{C_{2,t}^{i}}$, and $\widehat{C_{3,t}^{i}}$,  only $\widehat{C_{3,t}^{i}}$ depends on the latest estimation of $\theta$, since $\widehat{C_{3,t}^{i}}= 1_{\{M^{i}=0\}}\left(1-e^{-\widehat{\theta^{(t)}}U_{i}}\right)=1_{\{M^{i}=0\}}\left(1-e^{-{e^{-(\widehat{\beta_0^{(t)}}+\widehat{\beta^{(t)}}^T\textbf{X}_i)}}U_{i}}\right)$. The M-Step of the new algorithm  estimates $\gamma$ and $q$, as well as $\beta_{0}$ to $\beta_{k}$. 

We calculate the partial derivative  of \eqref{eq:log_lik_cov_raw} with respect to $\beta_{j}$ as follows:

\[
\begin{aligned}\frac{\partial\ell}{\partial\beta_{j}} & =\sum_{i=1}^{n}\left(\widehat{C_{1,t}^{i}}\right)\left(X_{i,j}U_{i}e^{-(\beta_0+\beta^T\textbf{X}_i)}\right)+\sum_{i=1}^{n}\left(\widehat{C_{2,t}^{i}}\right)(-X_{i,j})\left(1-U_{i}e^{-(\beta_0+\beta^T\textbf{X}_i)}\right)\\
 & +\sum_{i=1}^{n}\left(\widehat{C_{3,t}^{i}}\right)
 \frac{X_{i,j}U_{i}e^{-(\beta_0+\beta^T\textbf{X}_i)}\left(-e^{-U_{i}e^{-(\beta_{0}+\beta^T\textbf{X}_{i})}}\right)} 
 {1-e^{-U_{i}e^{-(\beta_{0}+\beta^T\textbf{X}_{i})}}}. 
\end{aligned}
\]

Let $G_i$ denote $e^{-U_{i}e^{-(\beta_{0}+\beta^T\textbf{X}_{i})}}$.

\[
\begin{aligned}\frac{\partial\ell}{\partial\beta_{j}} 
 & =\sum_{i=1}^{n} \left[ \left(\widehat{C_{1,t}^{i}}\right)\left(-X_{i,j}\log{G_{i}}\right)+\left(\widehat{C_{2,t}^{i}}\right)(-X_{i,j})\left(1+\log{G_{i}}\right) +\left(\widehat{C_{3,t}^{i}}\right)\frac{X_{i,j} G_i \log{G_i} }{1-G_i} \right].
\end{aligned}
\]

Equating the partial derivatives above to zero, we get that the estimators for $\beta_{j}^{(t+1)}$ in the $t$th iteration of the algorithm are given as
the solution to the following $k+1$ estimating equations, where $j=0,...,k$:
\begin{equation}\label{eq:Betas}
\begin{aligned}0= & \sum_{i=1}^{n}\left(\widehat{C_{2,t}^{i}}\right)(-X_{i,j})\protect
+ \sum_{i=1}^{n} (-X_{i,j})\log{\widehat{G_{i}^{(t+1)}}} \left\{ \widehat{C_{1,t}^{i}} +  \widehat{C_{2,t}^{i}} - \frac{  \widehat{C_{3,t}^{i}}  \widehat{G_{i}^{(t+1)}}  }{1-\widehat{G_{i}^{(t+1)}}}\right\} 
\end{aligned}
\end{equation}

where $\widehat{\beta^{(t+1)}}\triangleq\begin{bmatrix}\widehat{\beta_{1}^{(t+1)}} & \cdots & \widehat{\beta_{k}^{(t+1)}}\end{bmatrix}$ and $\widehat{G_{i}^{(t+1)}} \triangleq e^{-U_{i}e^{-(\widehat{\beta_{0}^{(t+1)}}+\widehat{\beta^{(t+1)}}^T\textbf{X}_{i})}}$.

\vspace{0.3cm}

\begin{algorithm}
\SetAlgoLined
\KwResult{$\widehat{\beta^{(t+1)}_{0}}$,..., $\widehat{\beta^{(t+1)}_{k}}$, $\widehat{q^{(t+1)}}$ and $\widehat{\gamma^{(t+1)}}$.}

Initialization: For every customer $i$, use Eq.~\eqref{eq:ci} to calculate $\widehat{C_{1,0}^{i}}$ and $\widehat{C_{2,0}^{i}}$ and $\widehat{C_{3,0}^{i}}=\hat{\pi}_{i} 1_{\{M^{i}=0\}}$, where $\hat{\pi_{i}}\in\left[0,1\right]$ is chosen randomly. 

To obtain the starting
parameters, $(\widehat{\theta^{(1)}},\widehat{q^{(1)}},\widehat{\gamma^{(1)}})$, solve Equations \eqref{eq:PartialDTheta} and \eqref{eq:PartialDGammaq}, respectively. \\
\vspace{6pt}
 \While{ $ \mid\widehat{\theta^{(t)}}-\widehat{\theta^{(t+1)}}\mid+\mid\widehat{q^{(t)}}-\widehat{q^{(t+1)}}\mid+\mid\widehat{\gamma^{(t)}}-\widehat{\gamma^{(t+1)}}\mid>\epsilon$}{
  E-step: For all  $i \in \{1,...,n\}$ and $j \in \{1,2,3\}$ use Eq.~\eqref{eq:ci} to calculate $\widehat{C_{j,t}^{i}}$ given the observed data
   $D=\{(U_{i},Y_{i},\Delta_{i},X_{i,1},...,X_{i,k})\,,i=1,...,n\}$ and the current estimations of the parameters $(\widehat{\theta^{(t)}},\widehat{q^{(t)}},\widehat{\gamma^{(t)}})$.   \\

\vspace{3pt}
  M-step: Maximize to obtain $(\widehat{\beta^{(t+1)}_{0}},..., \widehat{\beta^{(t+1)}_{k}},\widehat{q^{(t+1)}},\widehat{\gamma^{(t+1)}})$.\
  That is, update the estimations of the parameters using Eqs.~\eqref{eq:Betas} and~\eqref{eq:PartialDGammaq}, respectively, then calculate $\widehat{\theta^{(t+1)}}=e^{\beta_{0}+{\widehat{\beta^{(t+1)}}}^T\textbf{X}}$.
 }
 \caption{The Extended EM Algorithm}
 \label{EMCV}
\end{algorithm}

\section{EM Algorithm Validation}
\label{app:validation}

\subsection{Accuracy}
\label{app:EMaccuracy}
As a first examination, we want to evaluate the accuracy of the estimations provided by the EM algorithm and to compare them with the accuracy of previous methods suggested in the literature: \cite{Mandelbaum2013Data} (Method 1) and \cite{Yefenof2018} (Method 2). For this purpose, we simulate data for $\tau$, $W$, and $Y$ with specific parameters $\theta$, $q$, and $\gamma$. We compute $\Delta$ from the realization of $\tau$ and $W$ according to its definition ($\Delta=1_{\{\tau \leq W \}}$). We then estimate  $\widehat{\theta}$, $\widehat{q}$, and $\widehat{\gamma}$ using the EM algorithm to evaluate accuracy. Hence, in this validation strategy, all the assumptions of the EM algorithm hold.

As mentioned, the EM algorithm can cope with missing data, but the other two methods cannot.
To use them for this comparison, we need to make certain assumptions on how they handle conversations in the uSab class ($M=0$). To apply  \cite{Yefenof2018}, we have three options of how to classify $M=0$ conversations: classifying them either as served (Sr) customers ($C_{1}=1$) (denoted M2-Sr) or as silent-abandonment customers ($C_{3}=1$) (denoted M2-Sab), or misclassifying them using the same error rate as the SVM model (\S \ref{sec:prob_abnd_WQ}) (denoted M2-SVM).  Here, we simulate the last option by correctly classifying 85\% of the Sab conversations and 76\% of the Sr1 conversations.
To apply the method of \cite{Mandelbaum2013Data}, we have two options of how to classify $M=0$ conversations, either as served customers ($C_{1}=1$) (denoted M1-Sr) or as Kab ($C_{2}=1$) (denoted M1-Ab), since this method cannot deal with left-censoring.

\subsubsection{Accuracy with Regard to the Estimation of $\theta$.}
\label{app:EMaccuracy-theta}
We generate 200 samples of 2,000 customer conversations using the  parameter combinations stated in Table \ref{tbl:ParametersAcc}. 
Most of the parameters are taken from \cite{Yefenof2018} (Chapter 6), namely $\theta=4$ and $\gamma=10$ customers per hour (i.e., $E[\tau]=15$ and $E[W]=6$ minutes). We set $q$ to be in the set $\{1,0.9,...,0.1\}$, resulting in a proportion of silent abandonments between 0\% and 26\%. To create higher proportions of Sab customers between 27\% and $44\%$, we need to reduce $\gamma$; we use $\gamma \in \{9,7,5,4\}$ to achieve those abandonment rates. Note that the setting where $\theta<\gamma$ is plausible, since \cite{Brown2005}  found that average customer patience in call centers is greater than average virtual wait time, $E[\tau]>E[W]$. This result has been confirmed to hold in other service environments by several empirical studies, such as \cite{Yefenof2018}, who obtained this result when analyzing data from an ED. All the parameter combinations we choose are designed to keep the simulation within the same $\theta$ less than $\gamma$ setting.

\begin{figure}[htb]
\centering
\subfigure[MSE of $\theta$]{
\includegraphics[width=0.35\textwidth]{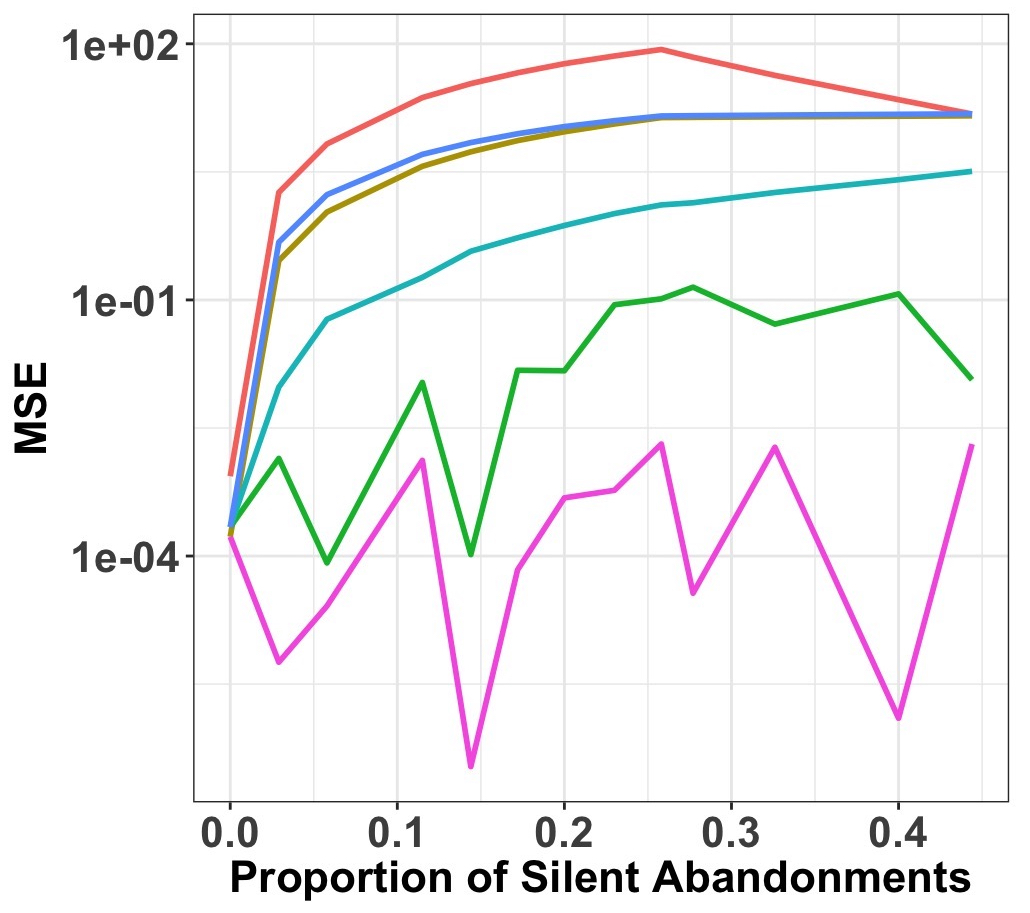} \label{fig:SimulatedData_a}
} 
\subfigure[MSE Ratio of $\theta$ (Baseline of the EM Algorithm)]{
\includegraphics[width=0.56\textwidth]{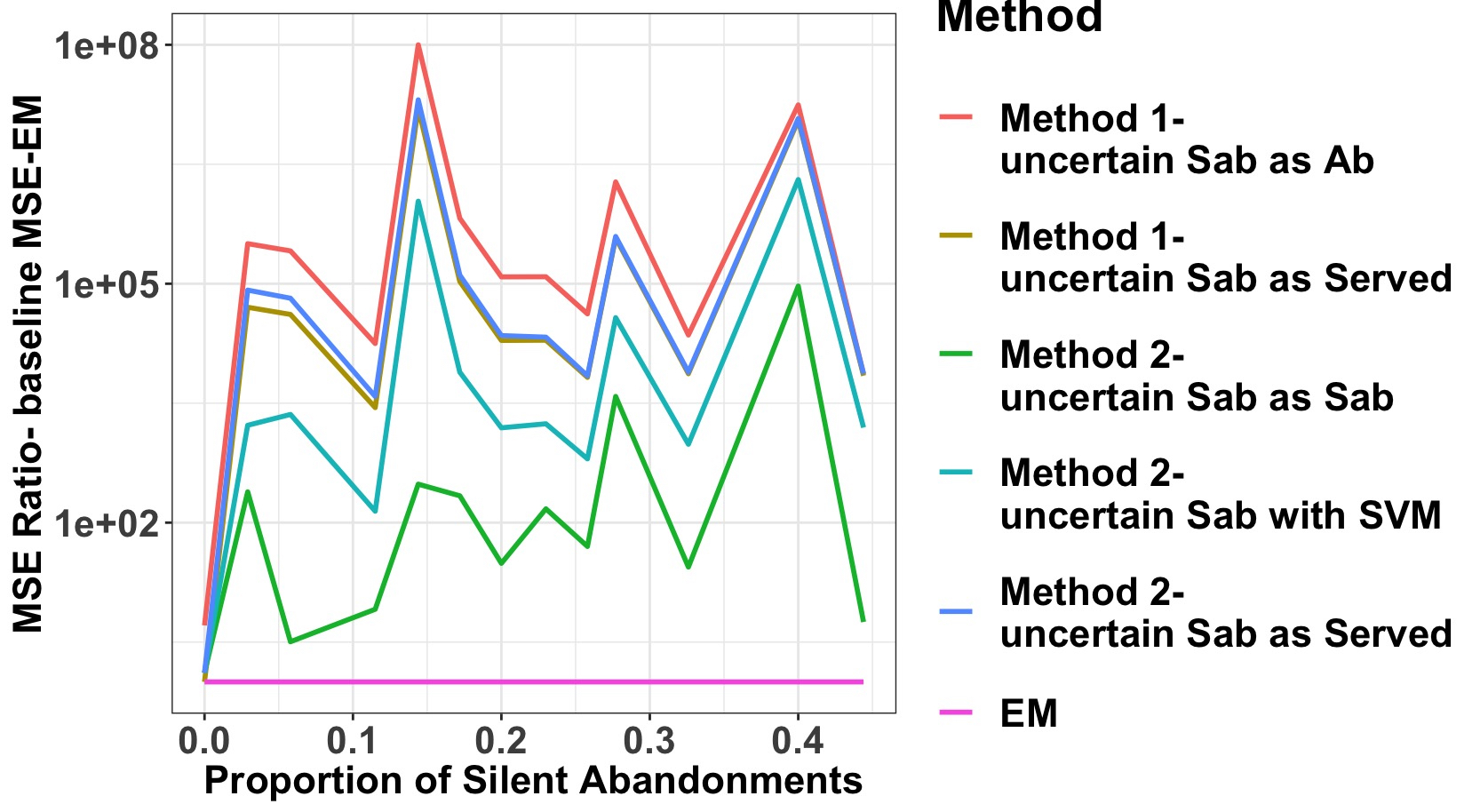}
\label{fig:SimulatedData_b}
}
\caption{Comparison of Accuracy of Customer-Patience Parameter, $\theta$, Estimations (Log Scale)}
\label{fig:SimulatedData}
\end{figure}

\begin{figure}[!htb]
\centering
\subfigure[2\% Sab ($\theta=4$, $\gamma=10$, $q=0.9$)]{
\includegraphics[width=0.31\textwidth]{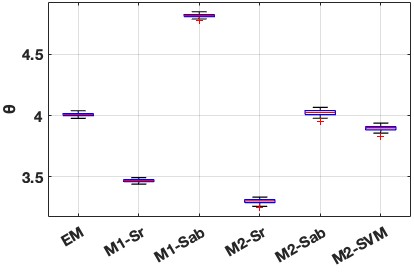} \label{fig:BoxSimulatedD_a}
} 
\subfigure[17\% Sab ($\theta=4$, $\gamma=10$, $q=0.4$)]{
\includegraphics[width=0.31\textwidth]{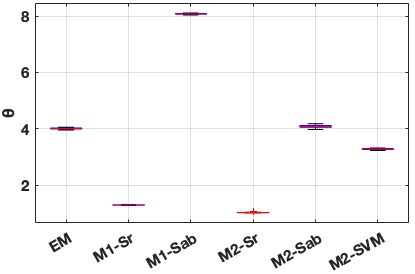} \label{fig:BoxSimulatedD_b}
} 
\subfigure[40\% Sab ($\theta=4$, $\gamma=5$, $q=0.1$)]{
\includegraphics[width=0.31\textwidth]{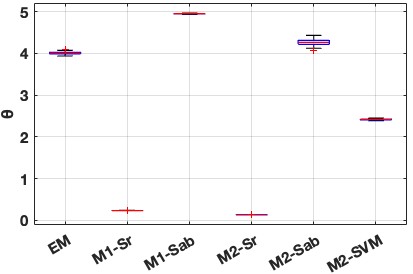} \label{fig:BoxSimulatedD_c}
} 
\caption{Accuracy of Customer-Patience Estimations for Low, Moderate, and High Sab Proportions}
\label{fig:BoxSimulatedD}
\end{figure}

For each sample, we estimate the parameters using the six methods mentioned above. 
We use 100 repetitions of the data sampling and parameter estimation with those methods to calculate the mean and SD of the estimated parameters (see Table \ref{tbl:ParametersAcc}) and create the boxplots (Figure \ref{fig:BoxSimulatedD}). 
Figure \ref{fig:SimulatedData} presents the accuracy results for estimating $\theta$ in a logarithmic scale. Figure EC.\ref{fig:SimulatedData_a} presents the mean squared errors (MSEs) for each model, while Figure EC.\ref{fig:SimulatedData_b} shows the ratio between the MSE of the specific model and the MSE of the EM algorithm (the baseline).
The x-axis in both figures is the proportion of silent abandonments of all arriving customers. Note that we do not report the results of any proportion of silent abandonments that is greater than 45\%, since we would not expect any company to find itself in such a position.

\begin{table}[!htb]
\centering
\caption{Real and Estimated Parameters in EM and Methods 1\&2 Accuracy Tests}
\begin{scriptsize}
\begin{tabular}{llllcrrrcrr} 
\toprule
P(Sab) & \multicolumn{3}{c}{Real} && \multicolumn{3}{c}{Expectation-Maximization} &&  \multicolumn{2}{c}{Method 1 - uSab as Sr} \\ \cline{2-4} \cline{6-8}  \cline{10-11}  
& $\theta$& $\gamma$& $q$ && \multicolumn{1}{c}{$\widehat{\theta}$}& \multicolumn{1}{c}{$\widehat{\gamma}$}& \multicolumn{1}{c}{$\widehat{q}$} && \multicolumn{1}{c}{$\widehat{\theta}$}& \multicolumn{1}{c}{$\widehat{\gamma}$} \tabularnewline
\midrule 
0 & 4& 10& 1.0&&4.01 [3.98, 4.03]&10.02 [9.99, 10.06] & 1.000 [1.000, 1.000] && 4.01 [3.95, 4.03] & 10.03 [9.99, 10.06] \\
0.03 & 4& 10& 0.9&& 4.00 [3.97, 4.02] & 10.00 [9.96, 10.03] & 0.899 [0.898, 0.900] && 3.31 [3.26, 3.33] & 10.00 [9.96, 10.03] \\
0.05 & 4& 10& 0.8&& 3.99 [3.97, 4.01] & 9.99 [9.96, 10.03] & 0.800 [0.798, 0.801] && 2.69 [2.65, 2.71] & 10.00 [9.96, 10.03] \\
0.08 & 4& 10& 0.7&& 4.03 [4.01, 4.05] & 9.98 [9.94, 10.01] & 0.598 [0.596, 0.601] && 1.75 [1.72, 1.76] & 9.98 [9.94, 10.01] \\
0.11 & 4& 10& 0.6&& 4.03 [4.01, 4.05] & 9.98 [9.94, 10.01] & 0.598 [0.596, 0.601] && 1.75 [1.72, 1.76] & 9.98 [9.94, 10.01] \\
0.14 & 4& 10& 0.5&& 4.00 [3.98, 4.01] & 9.99 [9.95, 10.02] & 0.499 [0.497, 0.502] && 1.36 [1.33, 1.37] & 9.99 [9.95, 10.02] \\
0.17 & 4& 10& 0.4&& 3.99 [3.98, 4.00] & 10.02 [9.99, 10.05] & 0.398 [0.396, 0.401] && 1.03 [1.00, 1.03] & 10.02 [9.99, 10.05] \\
0.2 & 4& 10& 0.3&& 4.02 [4.01, 4.03] & 10.00 [9.97, 10.03] & 0.300 [0.297, 0.302] && 0.73 [0.71, 0.73] & 10.00 [9.97, 10.03] \\
0.22 & 4& 10& 0.2&& 4.02 [4.01, 4.03] & 10.00 [9.96, 10.03] & 0.198 [0.196, 0.200] && 0.46 [0.44, 0.46] & 10.00 [9.96, 10.03] \\
0.25 & 4& 10& 0.1&& 4.04 [4.04, 4.05] & 10.00 [9.97, 10.03] & 0.100 [0.098, 0.102] && 0.22 [0.21, 0.22] & 10.00 [9.97, 10.03] \\
0.27 & 4& 9& 0.1&& 3.99 [3.99, 4.00] & 9.00 [8.97, ~9.03] & 0.101 [0.099, 0.102] && 0.21 [0.20, 0.21] & 9.01 [8.97,~ 9.03] \\
0.32 & 4& 7& 0.1&& 3.96 [3.95, 3.96] & 7.01 [6.98, ~7.03] & 0.100 [0.099, 0.102] && 0.18 [0.17, 0.18] & 7.01 [6.98, ~7.03] \\
0.4 & 4& 5& 0.1&& 4.00 [3.99, 4.01] & 4.99 [4.97, ~5.00] & 0.100 [0.098, 0.101] && 0.14 [0.13, 0.14] & 4.99 [4.97, ~5.00] \\
0.44 & 4& 4.1& 0.1&& 3.95 [3.95, 3.96] & 4.10 [4.09, ~4.11] & 0.099 [0.098, 0.100] && 0.11 [0.11, 0.11] & 4.10 [4.09, ~4.11] \\
\bottomrule
\multicolumn{11}{l}{Upper and lower 95\% confidence intervals inside brackets.}
\end{tabular}

\vspace*{5mm}

\begin{tabular}{llllcrrcrr} 
\toprule
P(Sab) & \multicolumn{3}{c}{Real} && \multicolumn{2}{c}{Method 1 - uSab as Ab} &&
\multicolumn{2}{c}{Method 2 - uSab as Sr}\\ \cline{2-4} \cline{6-7}  \cline{9-10}  
& $\theta$& $\gamma$& $q$ && \multicolumn{1}{c}{$\widehat{\theta}$}& \multicolumn{1}{c}{$\widehat{\gamma}$}&& \multicolumn{1}{c}{$\widehat{\theta}$}& \multicolumn{1}{c}{$\widehat{\gamma}$} \tabularnewline
\midrule 
0 & 4& 10& 1.0&& 4.01 [3.95, 4.03] & 10.03 [9.99, 10.06] && 4.01 [3.98, 4.03] & 10.03 [9.99, 10.06] \\
0.03 & 4& 10& 0.9&& 4.04 [3.98, 4.06] & 10.00 [9.96, 10.03] && 3.46 [3.43, 3.48] & 10.00 [9.96, 10.03] \\
0.05 & 4& 10& 0.8&& 3.99 [3.92, 4.01] & 10.00 [9.96, 10.03] && 2.96 [2.94, 2.98] & 10.00 [9.96, 10.03] \\
0.08 & 4& 10& 0.7&& 4.10 [4.00, 4.12] & 9.98 [9.94, 10.01] && 2.08 [2.06, 2.09] & 9.98 [9.94, 10.01] \\
0.11 & 4& 10& 0.6&& 4.10 [4.00, 4.12] & 9.98 [9.94, 10.01] && 2.08 [2.06, 2.09] & 9.98 [9.94, 10.01] \\
0.14 & 4& 10& 0.5&& 4.01 [3.89, 4.02] & 9.99 [9.95, 10.02] && 1.67 [1.65, 1.68] & 9.99 [9.95, 10.02] \\
0.17 & 4& 10& 0.4&& 4.12 [3.96, 4.12] & 10.02 [9.99, 10.05] && 1.29 [1.27, 1.30] & 10.02 [9.99, 10.05] \\
0.2 & 4& 10& 0.3&& 4.12 [3.93, 4.11] & 10.00 [9.97, 10.03] && 0.95 [0.93, 0.95] & 10.00 [9.97, 10.03] \\
0.22 & 4& 10& 0.2&& 4.30 [4.02, 4.25] & 10.00 [9.96, 10.03] && 0.61 [0.59, 0.61] & 10.00 [9.96, 10.03] \\
0.25 & 4& 10& 0.1&& 4.32 [3.85, 4.16] & 10.00 [9.97, 10.03] && 0.30 [0.29, 0.30] & 10.00 [9.97, 10.03] \\
0.27 & 4& 9& 0.1&& 4.38 [3.91, 4.22] & 9.01 [8.97, ~9.03] && 0.29 [0.28, 0.29] & 9.01 [8.97, ~9.03] \\
0.32 & 4& 7& 0.1&& 4.23 [3.83, 4.11] & 7.01 [6.98, ~7.03] && 0.27 [0.26, 0.27] & 7.01 [6.98, ~7.03] \\
0.4 & 4& 5& 0.1&& 4.34 [3.99, 4.26] & 4.99 [4.97, ~5.00] && 0.23 [0.23, 0.24] & 4.99 [4.97, ~5.00] \\
0.44 & 4& 4.1& 0.1&& 4.11 [3.85, 4.07] & 4.10 [4.09, ~4.11] && 0.21 [0.21, 0.21] & 4.10 [4.09, ~4.11] \\
\bottomrule
\multicolumn{9}{l}{Upper and lower 95\% confidence intervals inside brackets.}
\end{tabular}

\vspace*{5mm}

\begin{tabular}{llllcrrcrr} 
\toprule
P(Sab) & \multicolumn{3}{c}{Real} && \multicolumn{2}{c}{Method 2 - uSab as Sab} && \multicolumn{2}{c}{Method 2 - uSab with SVM}\\
\cline{2-4} \cline{6-7}  \cline{9-10} 
&\multicolumn{1}{c}{$\widehat{\theta}$}& \multicolumn{1}{c}{$\widehat{\gamma}$}&
\multicolumn{1}{c}{$\widehat{q}$}&&  \multicolumn{1}{c}{$\widehat{\theta}$}& \multicolumn{1}{c}{$\widehat{\gamma}$} && \multicolumn{1}{c}{$\widehat{\theta}$}& \multicolumn{1}{c}{$\widehat{\gamma}$}\tabularnewline
\midrule 
0 & 4& 10& 1.0&& 4.01 [3.98, 4.03] & 10.03 [9.99, 10.06] && 4.01 [3.95, 4.03] & 10.03 [9.99, 10.06] \\
0.03 & 4& 10& 0.9&& 4.81 [4.78, 4.83] & 8.65 [8.61, 8.68] && 3.90 [3.85, 3.92] & 10.00 [9.96, 10.03] \\
0.05 & 4& 10& 0.8&& 5.56 [5.52, 5.58] & 7.41 [7.37, 7.43] && 3.76 [3.70, 3.77] & 10.00 [9.96, 10.03] \\
0.08 & 4& 10& 0.7&& 6.91 [6.87, 6.93] & 5.16 [5.13, 5.18] && 3.57 [3.51, 3.59] & 9.98 [9.94, 10.01] \\
0.11 & 4& 10& 0.6&& 6.91 [6.87, 6.93] & 5.16 [5.13, 5.18] && 3.57 [3.51, 3.59] & 9.98 [9.94, 10.01] \\
0.14 & 4& 10& 0.5&& 7.51 [7.48, 7.53] & 4.15 [4.12, 4.17] && 3.39 [3.32, 3.41] & 9.99 [9.95, 10.02] \\
0.17 & 4& 10& 0.4&& 8.07 [8.04, 8.10] & 3.24 [3.21, 3.25] && 3.27 [3.21, 3.28] & 10.02 [9.99, 10.05] \\
0.2 & 4& 10& 0.3&& 8.60 [8.57, 8.62] & 2.35 [2.33, 2.36] && 3.14 [3.08, 3.15] & 10.00 [9.97, 10.03] \\
0.22 & 4& 10& 0.2&& 9.10 [9.06, 9.12] & 1.52 [1.50, 1.52] && 2.99 [2.92, 3.00] & 10.00 [9.96, 10.03] \\
0.25 & 4& 10& 0.1&& 9.57 [9.53, 9.59] & 0.74 [0.72, 0.74] && 2.86 [2.79, 2.87] & 10.00 [9.97, 10.03] \\
0.27 & 4& 9& 0.1&& 8.64 [8.61, 8.66] & 0.65 [0.64, 0.66] && 2.83 [2.76, 2.84] & 9.01 [8.97, ~9.03] \\
0.32 & 4& 7& 0.1&& 6.81 [6.78, 6.83] & 0.47 [0.46, 0.47] && 2.65 [2.60, 2.66] & 7.01 [6.98, ~7.03] \\
0.4 & 4& 5& 0.1&& 4.93 [4.91, 4.94] & 0.29 [0.29, 0.29] && 2.40 [2.36, 2.41] & 4.99 [4.97, ~5.00] \\
0.44 & 4& 4.1& 0.1&& 4.09 [4.08, 4.10] & 0.22 [0.22, 0.22] && 2.21 [2.17, 2.22] & 4.10 [4.09, ~4.11] \\
\bottomrule
\multicolumn{9}{l}{Upper and lower 95\% confidence intervals inside brackets.}
\end{tabular}
\end{scriptsize}
\label{tbl:ParametersAcc}
\end{table}

Table \ref{tbl:ParametersAcc} and Figure EC.\ref{fig:SimulatedData_a} show that the errors of the EM algorithm are quite small (less than 0.2\%) in all of the parameter combinations.
Figure EC.\ref{fig:SimulatedData_b} shows that both ways of implementing Method 1 (which accounts only for right-censored data) are very inaccurate. Specifically, estimating customer patience while ignoring the silent-abandonment phenomenon altogether results in an error rate that is $O(10^{8})$ higher than that of the EM baseline. A similar picture emerges when implementing Method 2, which assumes that all the uncertain conversations are served. Here too, the error rate is $O(10^{8})$ higher than that of the EM algorithm. 
If we take silent-abandonment conversations into account to the extent that we regard them as left-censored conversations but ignore the missing data, we obtain a (relatively) lower error rate. This is apparent when we look at the other two ways of implementing Method 2: either by considering all missing data to be Sab conversations or by completing the data with an SVM model. The problem with the latter approach is that the classification is considered correct, whereas a classification model is not completely accurate but has certain  sensitivity and specificity proportions. However, both of the abovementioned options yield less accurate results than the EM algorithm does: the respective error rates are $O(10^{5})$ and $O(10^{7})$ greater than that of the EM algorithm. To conclude, our algorithm outperforms all other methods for estimating customer patience. Note that when there is no silent abandonment in the system (0\% in Figure \ref{fig:SimulatedData}), all methods achieve the same performance level; this suggests that the EM algorithm can be used also in cases where the company does not have Sab customers or is unsure whether they exist. 


In order to analyze whether the estimations are biased or just have larger variance, we present the boxplots in Figure \ref{fig:BoxSimulatedD}. We include boxplots for only three of the parameter combinations we simulated. The parameters were chosen to enable comparison of estimations of parameters that result in low (2\%), moderate (17\%), and high (40\%) levels of silent abandonment in subfigures (a), (b), and (c), respectively. We see that regardless of the level of silent abandonment, the EM algorithm produces the most accurate estimation of customer patience, followed by Method 2 taking uSab as Sab (M2-Sab), which overestimates $\theta$ (underestimates average customer patience). 


\subsubsection{Accuracy with Regard to the Estimation of $q$ and $\gamma$.}
\label{app:EMaccuracy-QandGamma}

\begin{figure}[htb]
\centering
\subfigure[MSE of $q$]{
\includegraphics[width=0.355\textwidth]{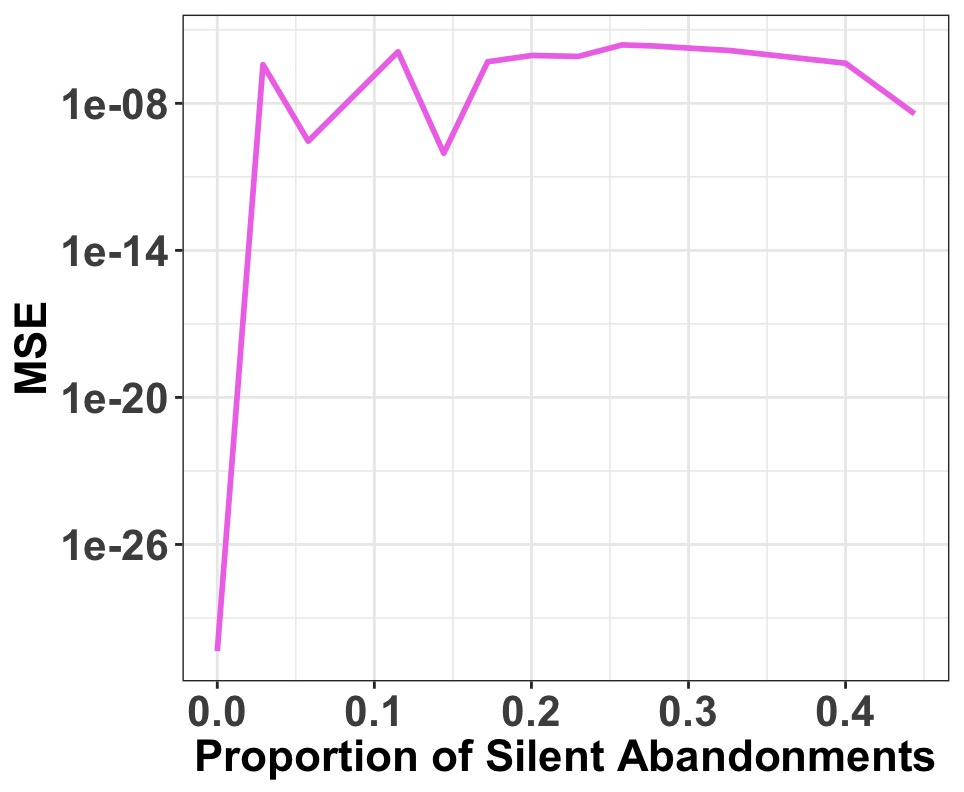} \label{fig:SimulatedDataq_a}
} 
\subfigure[MSE of $\gamma$]{
\includegraphics[width=0.555\textwidth]{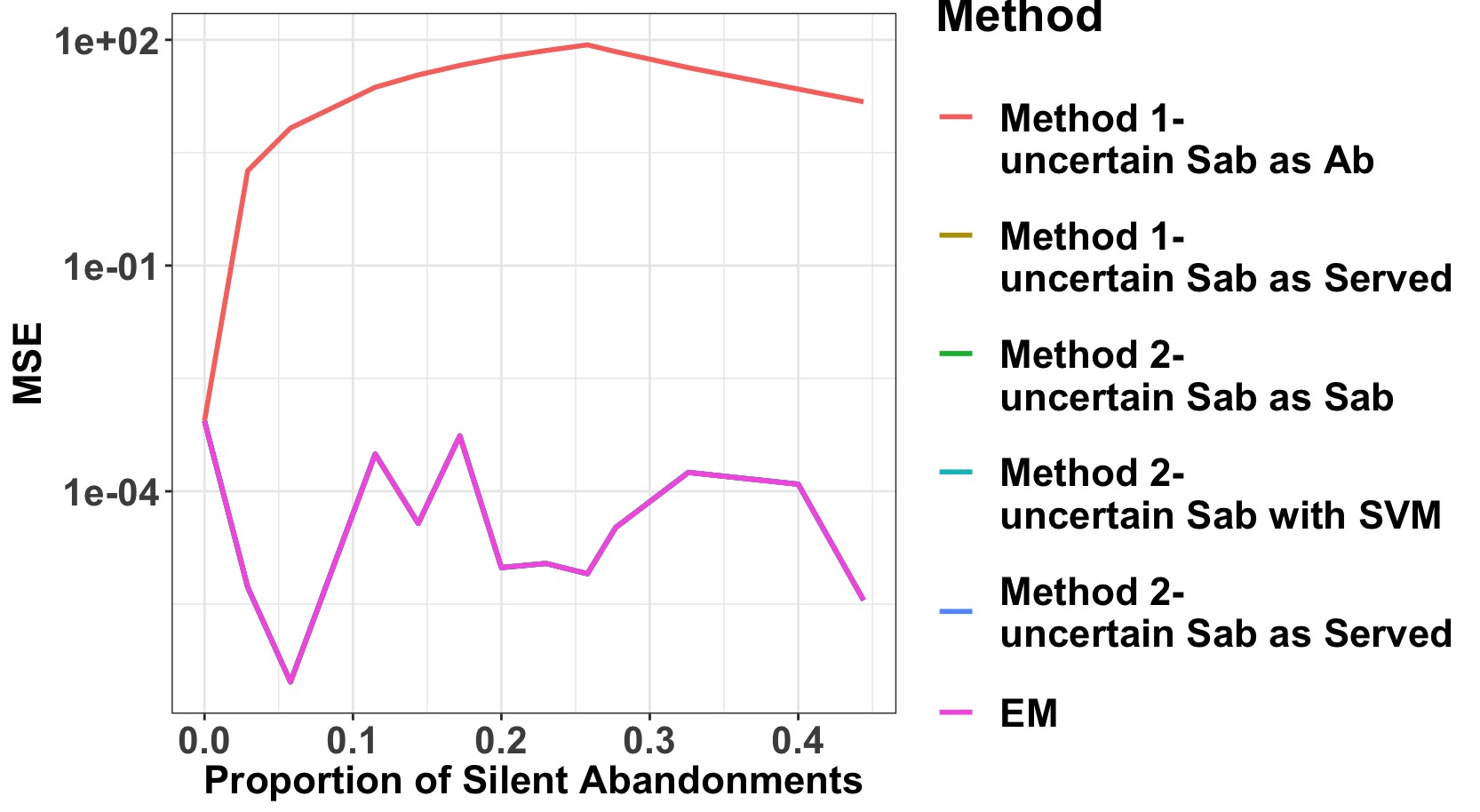}
\label{fig:SimulatedDataGamma_b}
}
\caption{Comparison of Accuracy of Probability of Indicating Abandonment ($q$) and Virtual Wait Time Parameter ($\gamma$), Estimations (Log Scale). Note That the $\gamma$ Estimation of All the Methods (Except One) Is Exactly the Same as for the EM Algorithm}
\label{fig:SimulatedDataGamma}
\end{figure}

Here, we provide the MSEs of the estimations of $q$ and $\gamma$ using the simulated data described in Appendix \ref{app:EMaccuracy}. Note that here we cannot compare estimation of $q$ to other methods since only the EM algorithm estimates this parameter. Figure EC.\ref{fig:SimulatedDataq_a} presents the MSE of $q$ as a function of the proportion of silent-abandonment customers (out of all the arriving customers) in logarithmic scale. We show here that the error rate is very small; therefore, the estimation is very accurate. 


Figure EC.\ref{fig:SimulatedDataGamma_b} presents the MSE results for estimating $\gamma$ in a logarithmic scale. 
The x-axis is the proportion of silent-abandonment customers (out of all the arriving customers). We note that the estimation of most of the methods is exactly the same, except for the method of \cite{Mandelbaum2013Data}, where we take the uncertain conversations ($M=0$) to be $C_{2}$. For this reason, most of the lines in Figure EC.\ref{fig:SimulatedDataGamma_b} are exactly the same as in the estimation of the EM algorithm. We conclude that the error rate of the EM algorithm as well as for most of the other methods  is quite small. 

\subsection{Sensitivity Analysis} \label{app:EMSensitivity}
The next tests are designed to investigate the sensitivity of the EM algorithm under the initial conditions. In addition, we would like to know whether starting the algorithm under some sophisticated initial conditions, for example by using a classification model, such as the one we developed in Section \ref{sec:prob_abnd_WQ}, helps the model converge to a more accurate estimation. Accordingly, we first investigate the sensitivity of the EM algorithm to $\hat{\pi}_{i}$. Note that by Algorithm \ref{EM}, $\hat{\pi}_{i}$  affects $\widehat{C_{3,0}^{i}}$ and $\widehat{C_{1,0}^{i}}$ only for the class of uSab customers.  

We generated 200 samples of 2,000 customer conversations using the following parameters: $\theta=4, \gamma=10$
, and $q=0.5$. For each sample, we estimate the parameters ($\widehat{\theta}, \widehat{q}$, $\widehat{\gamma}$) using the EM algorithm (with 100 repetitions) and consider the average of those parameters as the final estimator for that sample.
We present here four variants of the starting weights for all $M^{i}=0$ conversations.
\begin{lyxlist}{00.00.0000}
\item[\emph{All Sab:}] Setting all $M^{i}=0$ conversations to be silent-abandonment conversations with probability 1. Formally,  $\widehat{C_{3,0}^{i}}=1$ and $\widehat{C_{1,0}^{i}}=0$ for all conversations with $M^{i}=0$.   
\item[\emph{All Sr:}] Setting all $M^{i}=0$ conversations to be served-in-one-exchange conversations with probability 1. Formally,  $\widehat{C_{3,0}^{i}}=0$ and $\widehat{C_{1,0}^{i}}=1$ for all conversations with $M^{i}=0$.
\item[\emph{50:50:}] Setting 50\% of the conversations to be short-service conversations and 50\% to be Sab conversations; that is, for 50\% of the conversations with $M^{i}=0$, we set $\widehat{C_{3,0}^{i}}=1$, and for the rest of the $M^{i}=0$ conversations, we set  $\widehat{C_{1,0}^{i}}=1$. We choose this option because about 50\% of the uSab conversations are Sab and about 50\% are Sr1 within our data (see~\S\ref{sec:prob_abnd_WQ}).
\item [\emph{Best classifier:}] For conversations with $M^{i}=0$, we simulate a classification with sensitivity and specificity proportions according to our best classification model from Section \ref{sec:prob_abnd_WQ}; therefore, 85\% of the Sab conversations are classified correctly, as are 76\% of the Sr1 conversations. That is, 85\% of the actual $C_{3}=1$ conversations and 76\% of the actual $C_{1}=1$ conversations are identified as such; therefore, we set the correct values to $\widehat{C_{3,0}^{i}}$ and $\widehat{C_{1,0}^{i}}$ for those. 
For the remainder of the conversations, we set wrong values on  $\widehat{C_{3,0}^{i}}=1$ and $\widehat{C_{1,0}^{i}}=1$; for example,
for an actual $C_{3}=1$: $\widehat{C_{3,0}^{i}}=0$, $\widehat{C_{1,0}^{i}}=1$. 

\end{lyxlist}

\begin{figure}[!htb]
\centering
\includegraphics[width=0.3\textwidth]{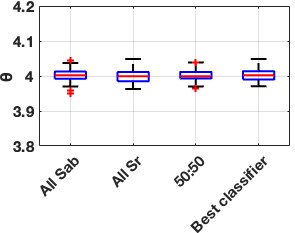}
\caption{Sensibility Analysis (Setting: $\theta=4$)}
\label{fig:Sensitivity}
\end{figure}

Figure \ref{fig:Sensitivity} shows that the estimations of customer patience are stable and do not change when different initial values are inserted in the EM algorithm. This suggests that one may not need to use the output of the classification model we developed in Section \ref{sec:prob_abnd_WQ}  (or any model with similar sensitivity and specificity proportions) as starting probabilities in the EM algorithm. 



\begin{figure}[!htb]
\centering
\subfigure[$q=0.5$]{
\includegraphics[width=0.3\textwidth]{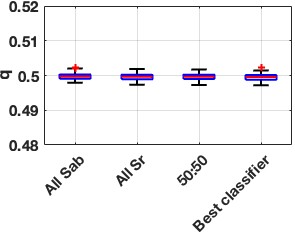} \label{fig:Sensitivityq}
} 
\subfigure[$\gamma=10$]{
\includegraphics[width=0.3\textwidth]{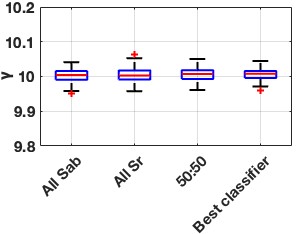} \label{fig:SensitivityGamma}
} 
\caption{Sensitivity Analysis}
\label{fig:More_Sensitivity}
\end{figure}

Next, we provide a sensitivity estimation for $q$ and $\gamma$ (see Figures EC.\ref{fig:Sensitivityq} and EC.\ref{fig:SensitivityGamma}, respectively),  using the same simulated data. The results are consistent, showing that the EM algorithm is not sensitive to the initial values.

\subsection{Robustness Check of Parameter Estimation of Real Data}
\label{app:EMRobustness}

A potential problem with EM algorithms is that they might converge to a saddle point (Chapter 8 of \citealt{little2002statistical}). 
To verify that this does not happen here, we started our EM algorithm with different weights. Specifically, we estimated the parameters by using the EM algorithm and taking the starting weights $\widehat{C_{3,0}^{i}}$ for the $M^{i}=0$ conversations to be $1$, $0$, $0.5$, or $\hat{\pi_{i}}$  from the SVM model, 
presented in Section \ref{sec:prob_abnd_WQ}. 
In every case, the obtained parameters ($\widehat{\theta}, \widehat{q}$, $\widehat{\gamma}$) were consistent, verifying that the algorithm did  not converge to a saddle point when applied to the real  data. 

\begin{figure}[!htb]
\centering
\includegraphics[width=0.4\textwidth]{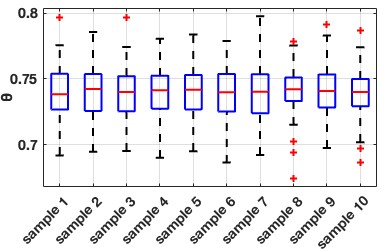}
\caption{Estimations of the Parameter of Patience ($\theta$) Using the EM Algorithm in Subsamples of the Write-in-queue Dataset (May 2017). Horizontal Line Indicates Estimation Based on the Entire Dataset $\hat{\theta}=0.739$}
\label{fig:tetaSamples}
\end{figure}

\begin{figure}[!htb]
\centering
\subfigure[Probability of Indicating Abandonment ($q$)]{
\includegraphics[width=0.4\textwidth]{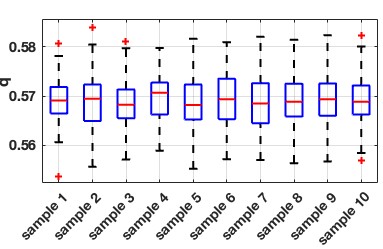} \label{fig:qSamples}
} 
\subfigure[Virtual Wait Time Parameter  ($\gamma$)]{
\includegraphics[width=0.4\textwidth]{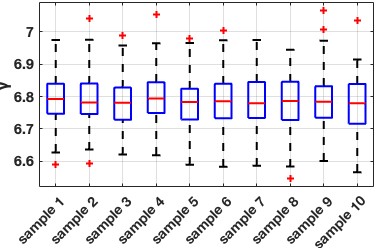} \label{fig:virtualWaitSamples}
} 
\caption{Estimated $q$ and $\gamma$ Using the EM Algorithm in Samples of the Write-in-queue Data, Horizontal Lines Indicate Estimation Based on the Entire Dataset $\hat{q}=0.58$ and $\hat{\gamma}=6.78$}
\label{fig:More_Samples}
\end{figure}

Finally, we performed several robustness checks by dividing the data set into 10--15 samples and estimating patience in each one using the EM algorithm 100 times. We performed these tests to make sure that the results we obtained from the monthly data ($\hat{\theta}=0.739$, $\hat{q}=0.58$, and $\hat{\gamma}=6.78$) are robust. We find that the estimations of $\theta$, $q$, and $\gamma$ from subsamples of the dataset are consistent with those of the whole dataset corpus. See Figures \ref{fig:tetaSamples}, EC.\ref{fig:qSamples}, and EC.\ref{fig:virtualWaitSamples}. 
%
%
Note that the estimation of $q$ using a smaller sample results in a small bias. 


\section{Mathematical Formulations for Method 1 \citep{Mandelbaum2013Data} and Method 2 \citep{Yefenof2018}}
\label{app:mandelbaum_yefenot_formulas}
For your convenience, we provide here details on Method 1 and 2, used in Section \ref{sec:patience_estimate} and Appendix \ref{app:validation}.

\textbf{Method 1---\cite{Mandelbaum2013Data}:}
In this method, only two classes of customers exist: served and known abandonments. Let $Y_{i}$ be an indicator of whether the customer abandoned, and $U_{i}$ the observed time of the customer in the queue. \cite{Mandelbaum2013Data} suggest the following estimates:
\[
\hat{\theta}=\frac{\sum_{i=1}^{n}Y_{i}\triangle_{i}}{\sum_{i=1}^{n}U_{i}},\;\hat{\gamma}=\frac{\sum_{i=1}^{n}\left(1-\triangle_{i}\right)}{\sum_{i=1}^{n}U_{i}}.
\]
When applying this method in Section \ref{sec:patience_estimate} and Appendix \ref{app:validation}, we can assume that the uSab customers are either known abandonment (Method 1---Uncertain silent abandonment is abandonment) or served (Method 1---Uncertain silent abandonment is service). For the first option, $Y_i=1$ for Kab and uSab customers and 0 otherwise, and in the second option, $Y_i=1$ only for Kab customers.

\textbf{Method 2---\cite{Yefenof2018}:}
In this method, three classes of customers exist: served, known abandonments, and silent abandonments. Let $Y_{i}$ be an indicator of whether the customer abandoned, $U_{i}$ the observed time of the customer in the queue, and $\triangle_{i}$ an indicator of whether the customer was served. \cite{Yefenof2018} suggest the following estimates:
\[
\theta=\frac{\sum_{i=1}^{n}\left(1-\triangle_{i}\right)}{\sum_{i=1}^{n}U_{i}\left(1-\triangle_{i}\right)}-\frac{n-\sum_{i=1}^{n}Y_{i}\triangle_{i}}{\sum_{i=1}^{n}U_{i}},\;\gamma=\frac{n-\sum_{i=1}^{n}Y_{i}\triangle_{i}}{\sum_{i=1}^{n}U_{i}}.
\]

When applying this method in Section \ref{sec:patience_estimate} and Appendix \ref{app:validation}, we can assume that the uSab customers are either silent abandonment (Method 2—Uncertain silent abandonment is silent abandonment) or served (Method 2---Uncertain silent abandonment is service). For the first option, $Y_i=1$ for Kab and uSab customers and 0 otherwise, and in the second option, $Y_i=1$ only for Kab customers. Moreover, for the first option, $\triangle_{i}=1$ for served customers only, and for the second option, $\triangle_{i}=1$ for served and uSab customers.

\end{document}